\documentclass[a4paper,11pt]{article}

\pdfoutput=1

\newcommand{\den}{\rho(\vec{x},\theta,t)}
\newcommand{\denbar}{\bar{\rho}(\vec{x},\theta,t)}
\newcommand{\denp}{\rho(\vec{x},\theta^\prime,t)}
\newcommand{\denpbar}{\bar{\rho}(\vec{x},\theta^\prime,t)}
             
\usepackage{jcappub}
\usepackage{amsmath}
\usepackage{amsfonts}
\usepackage{amssymb}
\usepackage{mathtools}
\usepackage{graphics}
\usepackage{citesort}
\usepackage{graphicx}
 \usepackage{url}
 \usepackage{xcolor}
\usepackage{cancel}
\usepackage[normalem]{ulem}
\usepackage[utf8]{inputenc}

\usepackage{hyperref}

\title{Neutrino propagation hinders fast pairwise flavor conversions}

\author{Shashank Shalgar,}
\author{Ian Padilla-Gay,}
\author{and Irene Tamborra}
\affiliation{Niels Bohr International Academy and DARK, Niels Bohr Institute, University of Copenhagen, Blegdamsvej 17, 2100, Copenhagen, Denmark}

\emailAdd{shashank.shalgar@nbi.ku.dk}
\emailAdd{ian.padilla@nbi.ku.dk}
\emailAdd{tamborra@nbi.ku.dk}

\abstract{Neutrino flavor conversions may dramatically affect the inner working of compact astrophysical objects as well as the synthesis of the heavier elements. We present the first sophisticated  numerical solution of the neutrino flavor conversion within a  (2+1+1) dimensional setup: we  include the advective term in the neutrino equations of motion and track the flavor evolution in two spatial dimensions, one angular variable, and time. 
Notably, the advective term  hinders the development of  neutrino pairwise conversions, if the conditions favoring such conversions (i.e., crossings between the angular distributions of $\nu_e$ and $\bar\nu_e$ or a non-negligible flux of neutrinos traveling backward with respect to the main propagation direction) exist for time scales shorter than the typical time scale of the advective term.
As a consequence, fast pairwise conversions can only occur when the conditions favoring flavor conversions are self-sustained and global, such as the ones induced by the lepton emission self-sustained asymmetry (LESA) in core-collapse supernovae. 
Our work  highlights the major impact of the dynamical evolution of the neutrino field on the growth of flavor instabilities and the strong interplay between classical and quantum effects. Critical limitations of  the linear stability analysis, used to predict neutrino flavor instabilities, are also pointed out.  
 }

\begin{document}
\maketitle
\flushbottom

\section{Introduction}
\label{sec:intro}

In the interior of neutrino dense astrophysical environments, such as neutron star mergers and core-collapse supernovae (SNe), neutrinos experience a non-negligible potential due to the presence of other neutrinos. This potential is analogous to the one due to electrons in the well known Mikheev-Smirnov-Wolfenstein (MSW) effect~\cite{Wolfenstein:1977ue, Mikheev:1986gs}.
The neutrino-neutrino scattering gives rise to an extremely fascinating phenomenology,  inducing  non-linear effects in the neutrino equations of motion~\cite{Mirizzi:2015eza,Duan:2010bg}. 
Notably, as a result of the non-linear nature of the evolution equations,  the  flavor evolution of neutrinos with different momenta is correlated. 

The non-linearity of the neutrino equations of motion  in compact astrophysical objects makes the solution of the neutrino flavor evolution extremely challenging, even when unrealistic simplifying assumptions are made. One of the first successful self-consistent calculations of neutrino flavor evolution including neutrino-neutrino interactions has been  performed assuming spherical symmetry and instantaneous decoupling of all neutrino flavors at the same radius~\cite{Duan:2005cp, Duan:2006an, Duan:2006jv,Fogli:2007bk,Fogli:2008pt}, the so-called ``neutrino-bulb model.'' Despite being extremely simplified, the calculation of neutrino flavor evolution within the bulb model  still proves to be challenging. In fact, it requires  the numerical solution of several millions of differential equations that may take up to several hundred CPU-hours depending on the desired accuracy. 

Any relaxation of the assumptions of the neutrino-bulb model  makes the numerical solution of the flavor problem unfeasible.  However, semi-analytical techniques proved that the non-linear nature of the problem implies that  the bulb model provides  different  results  than the ones obtained when some of the assumptions of this model are relaxed~\cite{Banerjee:2011fj,Raffelt:2013rqa,Duan:2014gfa, Abbar:2015mca}.

In addition, since the decoupling of different flavors occurs at different radii and neutrinos undergo flavor-dependent interactions, the initial angular distributions are flavor dependent. In some circumstances, this can lead to coherent pairwise conversion of neutrinos~\cite{Sawyer:2008zs,Sawyer:2015dsa,Izaguirre:2016gsx}. Pairwise conversions are fast in the sense that their characteristic time scale  is proportional to the neutrino number density, instead than the typical neutrino vacuum frequency~\cite{Izaguirre:2016gsx}. 
Favorable conditions for fast flavor conversions may occur in the proximity of the  neutrino decoupling region~\cite{Sawyer:2015dsa,Izaguirre:2016gsx,Capozzi:2018clo,Dasgupta:2016dbv}. Therefore, fast pairwise conversions may have  important implications for the neutrino-driven explosion mechanism in SNe as well as the nucleosynthesis of heavy elements.

One of the conditions that has been identified as being relevant to the development of fast pairwise conversions is the existence of crossings in the electron neutrino lepton number (ELN) or  a non-negligible flux of neutrinos propagating in the backward direction~\cite{Izaguirre:2016gsx,Capozzi:2017gqd,Yi:2019hrp}. In a spherically symmetric SN, the occurrence of ELN crossings  in the proximity of the decoupling region requires a  sharp radial evolution of the baryon density, with electron neutrino and antineutrino number densities being comparable~\cite{Shalgar:2019kzy,Morinaga:2019wsv}. 
Moreover,   localized regions of ELN crossings may also occur in the early SN stages. It is not clear whether in this case the neutrino flavor evolution may affect the SN physics on a macroscopic scale, or whether there are fast neutrino conversions at all~\cite{Shalgar:2019kzy, Johns:2019izj}.
 In neutron star mergers, the occurrence of ELN crossings seems to be favored by the more complex geometry and the natural excess of $\bar\nu_e$'s over $\nu_e$'s~\cite{Wu:2017qpc,Wu:2017drk}. 

The major implications of  the eventual occurrence of fast pairwise conversions in compact astrophysical objects has triggered a remarkable  effort from the community to better grasp  this phenomenon~\cite{Tamborra:2017ubu,Shalgar:2019kzy,Abbar:2019zoq,Abbar:2018shq,Abbar:2017pkh,DelfanAzari:2019tez,Nagakura:2019sig,Morinaga:2019wsv,Azari:2019jvr}, but there is a long road ahead. 
In fact, one of the major complications is related to the numerical solution of this  problem with  high enough  spatial and angular resolution, as dictated by the high frequency imposed by the neutrino fast conversions.

Given the complications induced by the non-linear nature of the system, and due to the spontaneous breaking of spatial and axial symmetries~\cite{Duan:2014gfa, Raffelt:2013rqa}, a minimum of two spatial dimensions is required to properly grasp the physics of the system. We here present the first sophisticated (2+1+1) dimensional modeling of the fast pairwise conversions in compact astrophysical objects; we solve the equations of motion in two spatial dimensions, one angular variable, and time and we include the advective term in the equations of motion. Within a simplified framework mimicking patches of the  dense SN core, we explore the time evolution of the neutrino and antineutrino distributions in the presence of localized ELN crossings and of extended regions of ELN crossings similar to the ones induced by the lepton emission self-sustained asymmetry (LESA)~\cite{Tamborra:2014aua}. 

The main goal of our work is to investigate under which conditions flavor instabilities may grow within a dynamical system. To do that, we explore two different scenarios reproducing global and localized regions of ELN crossings similar to the ones found in  hydrodynamical simulations of SNe~\cite{Tamborra:2014aua,Morinaga:2019wsv,Abbar:2018shq,DelfanAzari:2019tez,Nagakura:2019sig}. We also introduce the ``instability parameter'' and generalize the criteria under which  fast pairwise conversions may occur. 

This paper is organized as follows.  In Sec.~\ref{sec:fast} we discuss the conditions favoring the occurrence of fast pairwise conversions proposed in the literature and adapt them to our (2+1+1) dimensional model. We also provide a generalization of the criteria leading to favorable conditions for fast conversions by introducing the ``instability parameter.'' In Sec.~\ref{sec:num},   we describe the setup of our (2+1+1) model and its numerical implementation. In Sec.~\ref{sec:conv}, for the first time, we explore the impact of the  advective term on the ELN evolution and on the growth of flavor instabilities. We then explore the flavor evolution  when one localized ELN excess occurs and in the presence of an extended stripe of ELN crossings.  Finally, an outlook of our work and conclusions are presented in Sec.~\ref{sec:conc}. 

\section{Fast pairwise neutrino flavor conversion}
\label{sec:fast}
The non-linearity induced by  the neutrino-neutrino interactions makes the flavor evolution strongly dependent on the geometry and the number of dimensions of the system. 
Since at least two spatial dimensions are required for exploring any eventual effect of the advective term on the neutrino angular distributions, we   explore the evolution of the (anti)neutrino angular distributions in time, within a two-dimensional (2D) box, and for one angular variable. 
  In this Section, we introduce the neutrino equations of motion in the 2D box and generalize the criteria leading to fast pairwise conversions by introducing the ``instability parameter.''  

\subsection{Equations of motion}
Our system consists of a 2D box with width and height given by $L_{x}$ and $L_{y}$ and periodic boundary conditions. 
For each point $(x,y)$ in  the box,  $2 \times 2$ density matrices describe  the neutrino and antineutrino fields, respectively,  at time $t$:
\begin{eqnarray}
\den = 
\begin{pmatrix}
\rho_{ee} & \rho_{ex}\\
\rho_{ex}^{*} & \rho_{xx} 
\end{pmatrix}\ \quad\ \mathrm{and}\ \quad\ 
\denbar  =
\begin{pmatrix}
\bar{\rho}_{ee} & \bar{\rho}_{ex}\\
\bar{\rho}_{ex}^{*} & \bar{\rho}_{xx} 
\end{pmatrix}\ .
\end{eqnarray}
The density matrix of neutrinos is normalized such that $\mathrm{tr}(\rho)=1$ and we fix the asymmetry between neutrinos and antineutrinos such that  $\mathrm{tr}(\bar\rho)= a$. 
For each point $(x,y)$ in the box, the (anti)neutrino field has a distribution in momentum. For the sake of simplicity and since we intend to focus on fast pairwise conversions, we assume all (anti)neutrinos have the same energy, and the momentum is only determined by the angle $\theta$ with respect to the $y-$axis.

The flavor evolution is  determined by the following equations of motion for neutrinos and antineutrinos: 
\begin{eqnarray}
i\left(\frac{\partial}{\partial t} + \vec{v}\cdot\vec{\nabla}\right) \den 
&=& [H(\theta),\den]\ ,
\label{eom1}
\\
i\left(\frac{\partial}{\partial t} + \vec{v}\cdot\vec{\nabla}\right) \denbar
&=& [\bar{H}(\theta),\denbar]\ .
\label{eom2}
\end{eqnarray}
The advective term, $\vec{v}\cdot\vec{\nabla}$, depends on the velocity of the (anti-)neutrino field $\vec{v}$. The latter  has modulus equal to the speed of light $c$ and is oriented along the direction of propagation. The Hamiltonian, $H$, consists of a vacuum term that depends on the neutrino mixing angle $\theta_V$ and the vacuum frequency $\omega$ (assumed to be identical for all neutrinos in our system for simplicity), a term describing the interactions of neutrinos with the matter background with $\lambda$ being the interaction strength, and a self-interaction term, see e.g.~\cite{Mirizzi:2015eza}: 
\begin{eqnarray}
\label{eq:H}
H(\theta) = 
\frac{\omega}{2} \left(\begin{matrix}-\cos 2\theta_V & \sin 2\theta_V\\ \sin 2\theta_V & \cos 2\theta_V\end{matrix}\right)+  \left(\begin{matrix}\lambda & 0\\ 0 & 0\end{matrix}\right)+ \mu \int d \theta^\prime \left[\denp -  \denpbar \right]  \left[1 - \cos(\theta - \theta^{\prime})\right]\ . \nonumber \\
\end{eqnarray}
The Hamiltonian of antineutrinos, $\bar{H}(\theta)$, is the same as $H(\theta)$ except for $\omega$ which is replaced by $-\omega$.
In the latter term on the right hand side of Eq.~\ref{eq:H},  $\mu$  represents the strength of  neutrino-neutrino interactions 
\begin{eqnarray}
\label{eq:mu}
\mu = 10^2\ \mathrm{km}^{-1}\ .
\end{eqnarray}
In order to track the flavor evolution numerically in a reasonable number of CPU hours, our benchmark value for $\mu$ corresponds to the typical neutrino-neutrino interaction strength at $\mathcal{O}(100~\mathrm{km})$ from the SN core during the accretion phase (see, e.g., Fig.~22 of Ref.~\cite{Mirizzi:2015eza}). Larger values of $\mu$, descriptive of the neutrino self-interaction strength  in the proximity of the neutrino decoupling region, would lead to the development of flavor conversions on scales smaller than what we discuss here without changing the overall conclusions, also for what concerns the impact of neutrino advection on the flavor evolution. 

Unless otherwise specified, in the following, we assume $\omega = 0.1$~km$^{-1}$ as typical of neutrinos in compact astrophysical objects, and $\theta_{V}=10^{-6}$ in order to effectively   ignore the matter term~\cite{EstebanPretel:2008ni}. The  default value adopted for the advective velocity is $c=1$ in natural units (corresponding to  $3\times 10^{5}$~km/sec). In addition, as we will discuss in Sec.~\ref{sec:conv}, we neglect the collision term.

Significant evolution in the number of (anti-)neutrinos can occur only if the off-diagonal term of the Hamiltonian is not very small compared to the diagonal term. In the case of  neutrino-neutrino interactions, the magnitude of the off-diagonal component of the Hamiltonian is a dynamical quantity, whose initial seed is set by the momentum distribution of the density matrices (i.e., our vacuum frequency $\omega$). For a given  initial angular and spatial distribution, whether there will be significant flavor evolution depends on the temporal evolution of the off-diagonal elements of the density matrices which are connected to the probability of flavor transition. If the off-diagonal components of the density matrices grow with time (i.e., a flavor instability occurs), they will eventually lead to a change in the diagonal components of the density matrices which are directly connected to the spatial and temporal  evolution of the number density of the different flavors. 
For the sake of simplicity and  without loss of generality, in what follows, we assume   $\rho_{xx} (t=0~\mathrm{s}) = \bar\rho_{xx}(t=0~\mathrm{s}) = 0$,  $a = 0.5$, and 
$L_x=L_y=20$~km. If all off-diagonal terms of the density matrices are zero at $t=0$~s, then they will remain zero in the absence of the linear  terms of the Hamiltonian (vacuum or matter term). 

\subsection{Instability parameter}
The rate of growth of the off-diagonal terms of the density matrices (and therefore of the flavor instability) can be estimated by using the linear stability analysis for  given initial conditions~\cite{Banerjee:2011fj,Izaguirre:2016gsx}. In particular, Ref.~\cite{Izaguirre:2016gsx} found that fast pairwise conversions may be induced by ELN crossings or in the presence of a non-negligible backward flux. In the case of ELN crossings, the growth of flavor instabilities may be affected by the depth of ELN crossings~\cite{Martin:2019gxb,Yi:2019hrp}.  

We here introduce the ``instability parameter''  that  depends on the shape of the angular distributions of $\nu_e$ and $\bar\nu_e$ and it is approximately proportional to the growth rate of the off-diagonal components of the density matrices:
\begin{eqnarray}
\zeta &=& \frac{I_{1}I_{2}}{(I_{1}+I_{2})^{2}} 
\label{zeta:def}
\end{eqnarray} 
with
\begin{eqnarray}
\label{I1:def}
I_{1} &=& \int_{0}^{2\pi} \Theta \left[\rho_{ee}(\theta)-\bar{\rho}_{ee}(\theta)\right] d\theta\\ 
I_{2} &=& \int_{0}^{2\pi} \Theta \left[\bar{\rho}_{ee}(\theta)-\rho_{ee}(\theta)\right] d\theta 
\label{I2:def}
\end{eqnarray} 
where  $\Theta$ is the Heaviside function, which vanishes when the argument is negative and is equal to the identity operator otherwise. It should be noted that, since the definition of $\zeta$ contains two powers of $I_{1,2}$ in the numerator and in the denominator, $\zeta$ is independent of the overall normalization of the density matrices. The $\zeta$ parameter is zero when there is no ELN crossing or when there is no backward flux, therefore it generalizes the criteria outlined in Ref.~\cite{Izaguirre:2016gsx}. 

\section{Neutrino flavor evolution in a two-dimensional box}
\label{sec:num}

We explore two configurations of our 2D box. The first  scenario corresponds to the case where neutrinos and antineutrinos are initially localized in a small region in our 2D box which should mimic the evolution of random fluctuations occurring within the  inner SN core. The second scenario consists of neutrinos and antineutrinos that are initially located along a stripe in the 2D box. This would mimic the evolution of flavor conversions in the case of an extended region of ELN crossings such as in the presence of LESA. In the following, we will introduce the numerical framework adopted to explore the flavor evolution in all these configurations.

\subsection{Model setup}\label{sec:setup}

\begin{figure}
\centering
\hspace{-5mm}\includegraphics[width=0.57\textwidth]{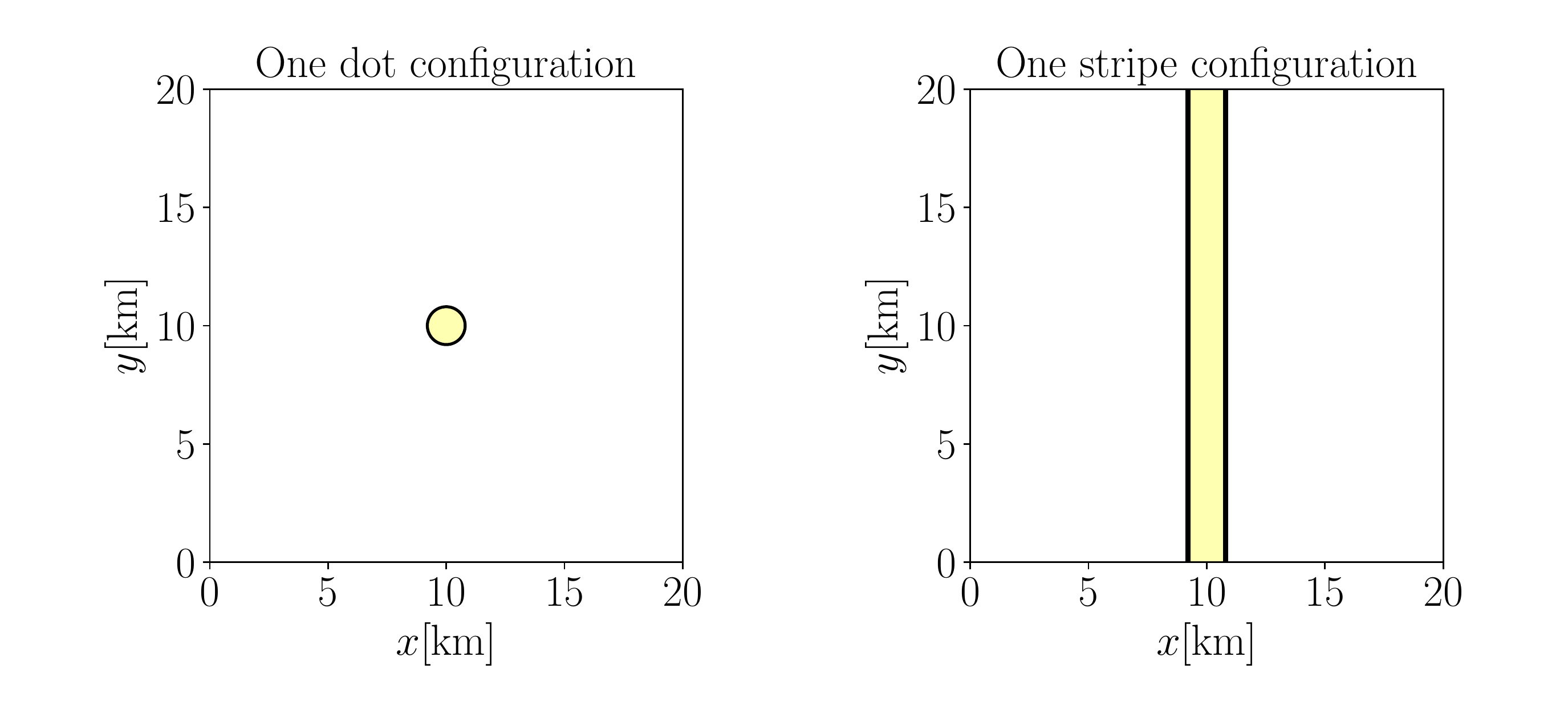}
\includegraphics[width=0.95\textwidth]{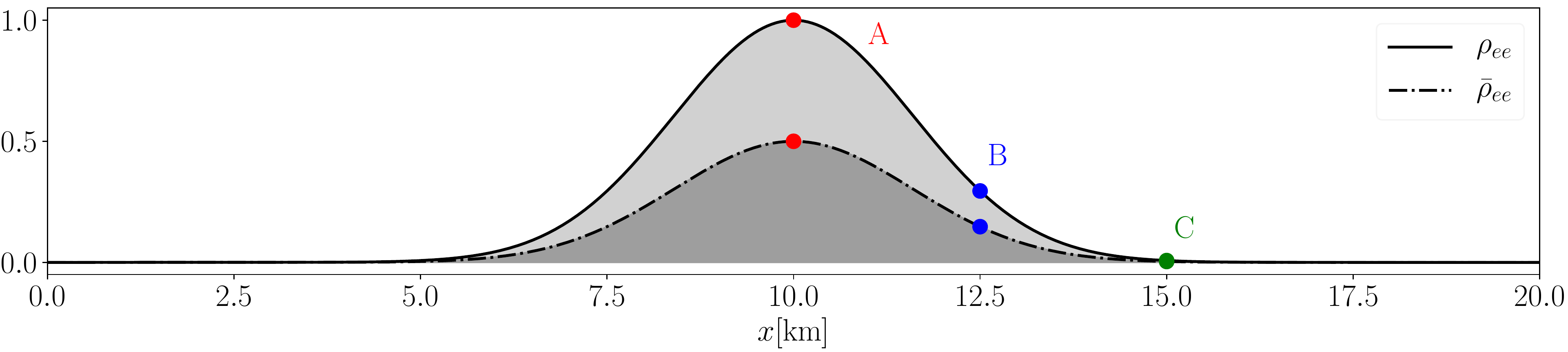}\\
\includegraphics[width=0.95\textwidth]{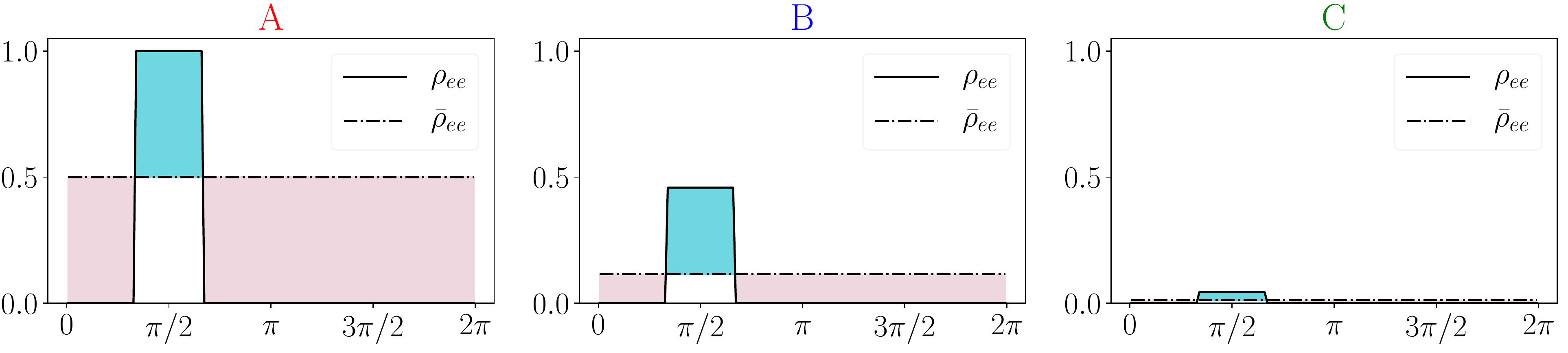}
\caption{{\it Top}: Sketch of the 2D box at $t=0$~s for the two configurations adopted in this paper: ``one dot configuration'' (on the left) and ``one stripe configuration'' (on the right),  see main text for more details. {\it Middle}: 1D projection of the dot or stripe  configuration to show $\rho_{ee}$ (continuous line) and $\bar\rho_{ee}$ (dot-dashed line) as a function of $x$. {\it Bottom}:  $\nu_e$ and $\bar\nu_e$ angular distributions for the three points $A$, $B$, and $C$ highlighted in the middle panel, respectively from left to right.}
\label{fig:1}
\end{figure}

We define a 2D  spatial grid with length $L_{x} = L_{y} = 20$~km which is identical in all simulations. 
The ``one dot configuration''  is shown in the  top panel on the left of Fig.~\ref{fig:1}. It has been obtained by assuming
\begin{eqnarray}
\label{eq:dot}
\rho_{ee}, \bar{\rho}_{ee} \propto \exp\left[-\frac{(x-x_{0})^{2}}{2\sigma^{2}}\right] \exp\left[-\frac{(y-y_{0})^{2}}{2\sigma^{2}}\right]\ ,
\end{eqnarray}
with each distribution centered on $x_0=y_0=1/2 L_x$ and $\sigma = 5\% L_x$. It corresponds to the initial box configuration  consisting of neutrinos distributed according to two Gaussians,  along the $x$($y$)-axis, each as displayed in the 1D projection in the middle panel of  Fig.~\ref{fig:1}. 

At $t=0$~s, for each $(x,y)$ the angular distributions of neutrinos and antineutrinos are fixed to be two top hat angular distributions centered on $\pi/2$,
\begin{eqnarray}
\rho_{ee}(\theta) = 
\begin{cases}
g & |\theta-\frac{\pi}{2}| < b \\
0 & |\theta-\frac{\pi}{2}| \ge b\ ,
\end{cases} 
\label{top1}
\\
\bar{\rho}_{ee}(\theta) = 
\begin{cases}
\bar{g} & |\theta-\frac{\pi}{2}| < \bar{b} \\
0 & |\theta-\frac{\pi}{2}| \ge \bar{b}\ ,
\end{cases}
\label{top2}
\end{eqnarray}
with $b$ being the opening angle of the $\nu_e$ angular distribution (assumed to be $\pi/6$ unless otherwise specified) and $\bar{b} = \pi$ being the one of $\bar\nu_e$.
This scenario aims to mimic the evolution of the ELN crossings generated by stochastic fluctuations in the proximity of the decoupling region. A sketch of the initial $\nu_e$ and $\bar\nu_e$ angular distributions for three selected points across $x$ is  shown in the bottom panel of Fig.~\ref{fig:1}.

The ``one stripe configuration''  is shown in the top  panel on the right of Fig.~\ref{fig:1}. It corresponds to the initial box configuration  consisting of neutrinos localized along a 
stripe, which is homogeneous along the $y$-axis and distributed according to a Gaussian along the $x$-axis as displayed in the 1D projection in the middle panel of  Fig.~\ref{fig:1}. The non-zero diagonal terms of the density matrix are defined as follows
\begin{eqnarray}
\label{eq:gauss}
\rho_{ee}, \bar{\rho}_{ee} \propto \exp\left[-\frac{(x-x_{0})^{2}}{2\sigma^{2}}\right]\ ,
\end{eqnarray}
with $\sigma = 5\% L_x$ and the center of the distribution $x_{0} = 1/2 L_x$.
This configuration mimics  what should happen in the presence of LESA, when the ELN changes its sign. 

Note that we assume that the angular distribution of $\nu_e$ is forward peaked and the $\bar\nu_e$ one is isotropic.  However, in a realistic framework, the angular distributions of neutrinos and antineutrinos are both forward peaked after decoupling.  As we will see later, we  focus on a more extreme scenario since  any growth of a flavor instability would be further suppressed, if both distributions are assumed to be forward peaked at $t=0$~s. Moreover, the occurrence of fast neutrino oscillations requires that the values of the heights ($g,\bar{g}$) and the widths ($b,\bar{b}$) of the angular distributions are different for neutrinos and antineutrinos.

\subsection{Numerical implementation}

We perform numerical simulations with different initial conditions, while keeping the overall architecture of the numerical simulations and the grid size unchanged. For each $(x,y)$ point, we define the angular distributions of neutrinos and antineutrinos as described in Sec.~\ref{sec:setup} and evolve the (anti)neutrino equation of motions according to Eqs.~\ref{eom1}, \ref{eom2}. In the numerical runs, we adopt the following number of spatial and angular bins: $N_x = N_y = 400$ and $N_\theta = 100$.  We use an adaptive method for the temporal evolution of the system. Note, however, that  for the cases without the advective term, the number of spatial bins is not relevant since the spatial gradient in the advective term is zero.  Although, we impose periodic  boundary conditions on our box, we let the stripe configuration evolve within a time interval such that (anti)neutrinos never cross the  boundaries.

The spatial gradient on the left-hand-side of Eqs.~\ref{eom1} and \ref{eom2} can be solved analytically in the absence of flavor evolution. This term is responsible for transporting the neutrino density matrices from one bin to another. To this purpose, we implement a transport algorithm to move the density matrices to neighbouring spatial bins in regular time steps.
For the temporal evolution we use the Runge-Kutta-Fehlberg(7,8) method from the odeint library of Boost~\cite{BoostLibrary}. 

In order to speed up the computational time, we parallelize our numerical code through the OpenMP interface~\cite{openmp}.  We evolve the simulations for $\mathcal{O}(10^{-5}~\mathrm{s})$ which is enough to gauge the flavor conversion phenomenology in astrophysical environments. Within this simplified setup, each simulation  run takes about 200 CPU-hours.

\subsection{Fast conversions in the absence of advection}
\begin{figure}
\centering
\includegraphics[width=0.9\textwidth]{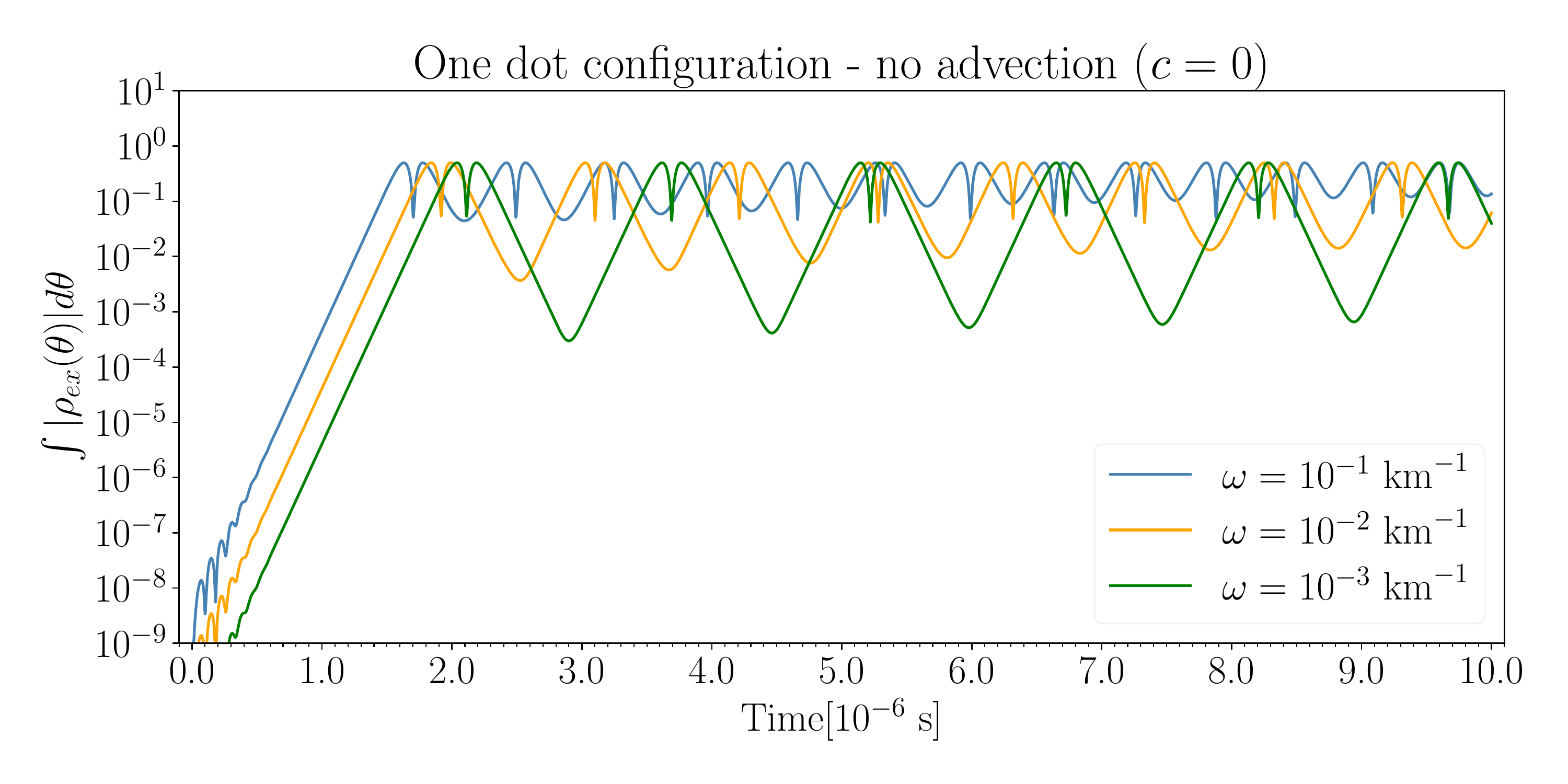}
\caption{Temporal evolution of $\int d\theta|\rho_{ex}(\theta)|$  in the absence of advection for the $(x_0,y_0)$ point in our 2D box and several values of vacuum frequency $\omega$. Flavor transformation occurs  when $\rho_{ex}\sim 1$ and gives rise to bipolar fast conversions. The oscillation  period is inversely correlated to $\omega$.
}
\label{fig:2}
\end{figure}

We now explore the general features of the linear and non-linear flavor evolution within our 2D box. 
Figure~\ref{fig:2}  shows an example of the outcome of our simulations  for the ``one dot configuration'' (top left of  Fig.~\ref{fig:1}). It shows the off-diagonal term of the density matrix $\int d\theta \rho_{ex}(\theta)$ as a function of time for the $(x_0,y_0)$ point in our 2D box in the absence of advection for $\omega=10^{-3}, 10^{-2}, 10^{-1}$~km$^{-1}$ (see Eq.~\ref{eq:H}).  As we will discuss in Sec.~\ref{sec:growth}, $\int d\theta \rho_{ex}(\theta)$ grows exponentially until it transitions to the non-linear regime.  When the magnitude of the off-diagonal elements of the density matrix reaches the same order as the diagonal elements ($\sim\rho_{ee}$),  flavor transformations start.  

Interestingly, in the non-linear regime, the neutrino flavor evolution closely resembles the bipolar oscillations commonly found for slow collective oscillations~\cite{Duan:2005cp,Hannestad:2006nj}. Although the growth rate at smaller times  is completely independent of the vacuum frequency, $\omega$, the non-linear evolution of  flavor is inversely correlated to $\omega$; in addition, a second frequency seem to affect the evolution of $\rho_{ex}$ independently of $\omega$. However, the latter does not affect the overall flavor evolution. This is a new finding for what concerns the phenomenology of fast pair-wise conversions. In fact, the vacuum term has been neglected in the stability analysis under the assumption that fast-pairwise conversions are completely driven by $\mu$, see e.g.~\cite{Izaguirre:2016gsx,Sawyer:2015dsa}, if crossings in the angular distributions of $\nu_e$ and $\bar\nu_e$ exist. However, we find that the separation in time between two bipolar transformations is inversely correlated to $\omega$. Although not shown here, we tested configurations with $\mu \in [10^2, 10^5]$~km$^{-1}$ as well as cases with $\lambda \neq 0$ and observed a  bipolar regime in all cases. As we will discuss in Sec.~\ref{sec:conv}, this picture will be modified by advection.

\subsection{Growth of the flavor instability}\label{sec:growth}
\begin{figure}
\begin{center}
\includegraphics[width=0.9\textwidth]{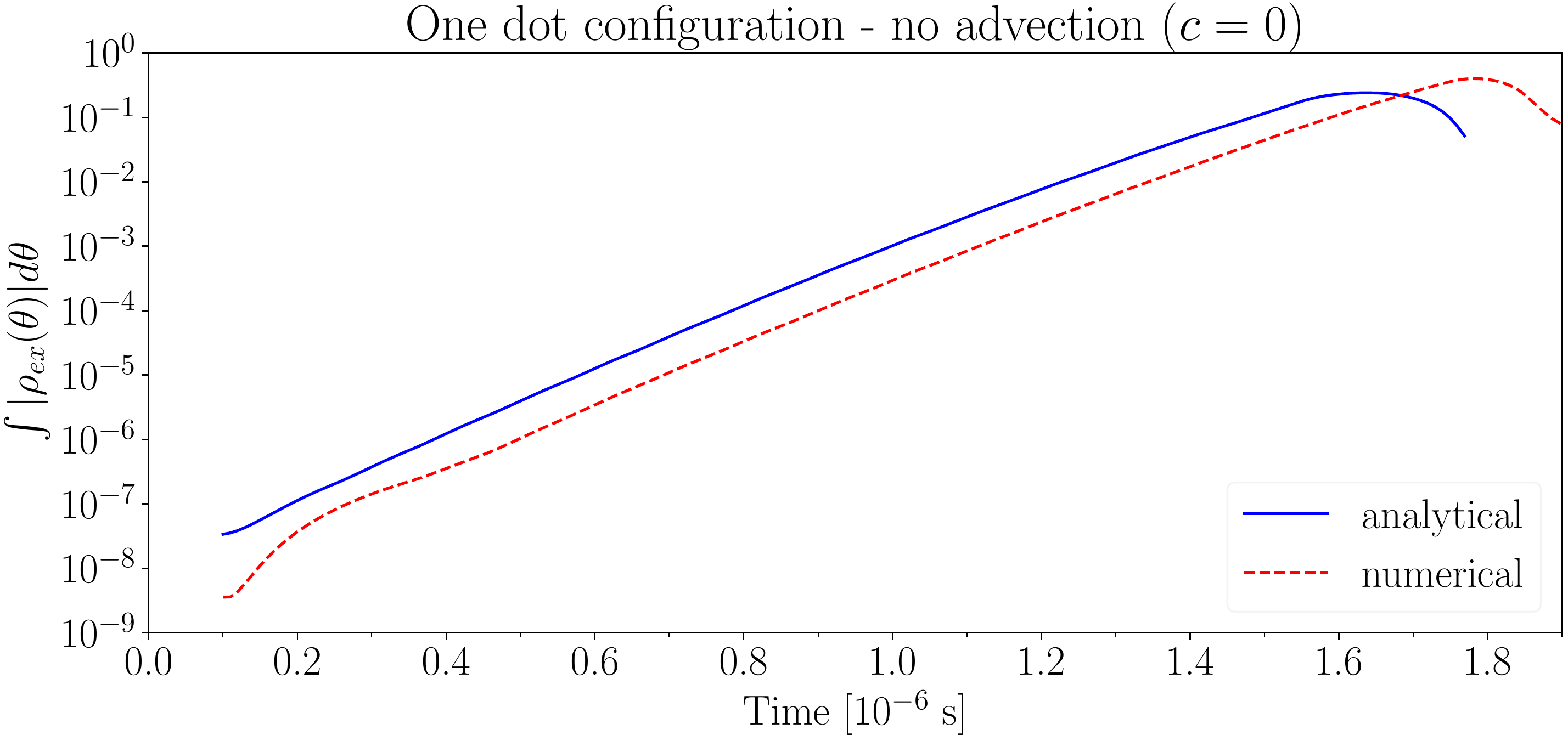}
\end{center}
\caption{Temporal evolution of $\int d\theta \rho_{ex}(\theta)$  for the $(x_0,y_0)$ point in the box (``one dot configuration'') in the absence of advection from our numerical simulation (dashed line) compared with the analytical approximation of Eq.~\ref{kappaprop} (continuous line).
The analytical expression of Eq.~\ref{kappaprop} perfectly matches the numerical growth rate of the instability.}
\label{fig:3}
\end{figure}
As shown in Fig.~\ref{fig:1}, when the advective term is neglected in Eqs.~\ref{eom1} and \ref{eom2}, the absolute value of the off-diagonal component of $\rho$ grows exponentially in time, before reaching the non-linear regime. We find that the exponent, $\kappa$,  satisfies the following relation
\begin{eqnarray}
\kappa = \epsilon  \mu_{\textrm{eff}} \zeta
\label{kappaprop}
\end{eqnarray}
with 
\begin{eqnarray}
\mu_{\mathrm{eff}} = \mu \int_{0}^{2\pi} d\theta \left[(\rho_{ee}(\theta)- \rho_{xx}(\theta)) - (\bar{\rho}_{ee}(\theta)- \bar{\rho}_{xx}(\theta))\right]\ ,
\end{eqnarray}
where this definition of $\mu_{\mathrm{eff}}$ depends on the ELN, while the one of $\mu$ in Eq.~\ref{eq:mu} is meant to be related to the total (anti)neutrino number density. The proportionality factor in Eq.~\ref{kappaprop} was found to be $\epsilon = 34.8$ for a wide range of initial configurations where $b(\bar{b})$ and $g(\bar{g})$ were varied according to Eqs.~\ref{top1} and~\ref{top2}.

The growth of the flavor instability is shown in Fig.~\ref{fig:3} as a dashed line as a function of time for the ``one dot configuration'' and for the $(x_0,y_0)$ point in the box. The continuous  line has been obtained by using Eq.~\ref{kappaprop}. As one can see, Eq.~\ref{kappaprop} perfectly reproduces the growth rate of the off-diagonal terms of the density matrix. 
It should be noted that the definition of $\kappa$ is heuristic in nature; although it is not clear whether a single parameter can encapsulate the growth rate of the instability for all angular distributions, the parametrization in Eq.~\ref{kappaprop} works for all cases that we have explored. 
When $\vec{v}\cdot\nabla = 0$, the exponential growth of the off-diagonal term of $\rho$ does not depend on the neighboring regions, and it continues to grow until the non-linear regime is reached.

\section{Role of the advective term in the neutrino evolution equation}
\label{sec:conv}
In this Section, after general considerations on the impact of the advective term, we discuss the role that the latter plays on the growth of flavor instabilities in the ``one dot configuration,'' 
and in the ``one stripe configuration.'' We also discuss how advection affects the evolution of the ELN distribution as a function of time. 

\subsection{Impact of the advective term on the neutrino distributions}\label{sec:arg}
\begin{figure}
\centering
\includegraphics[width=0.4\textwidth]{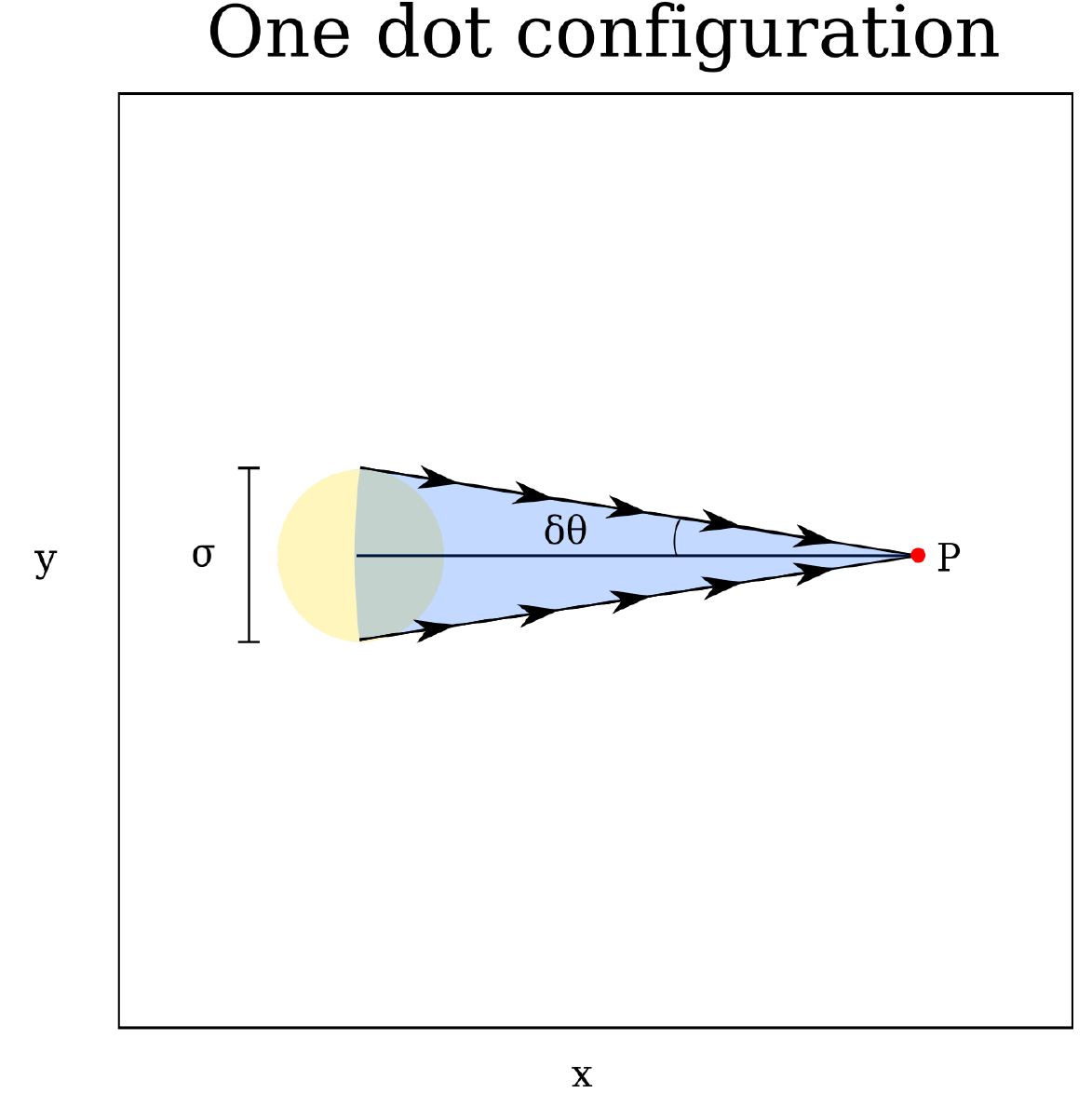}\hspace{0.5cm}
\includegraphics[width=0.395\textwidth]{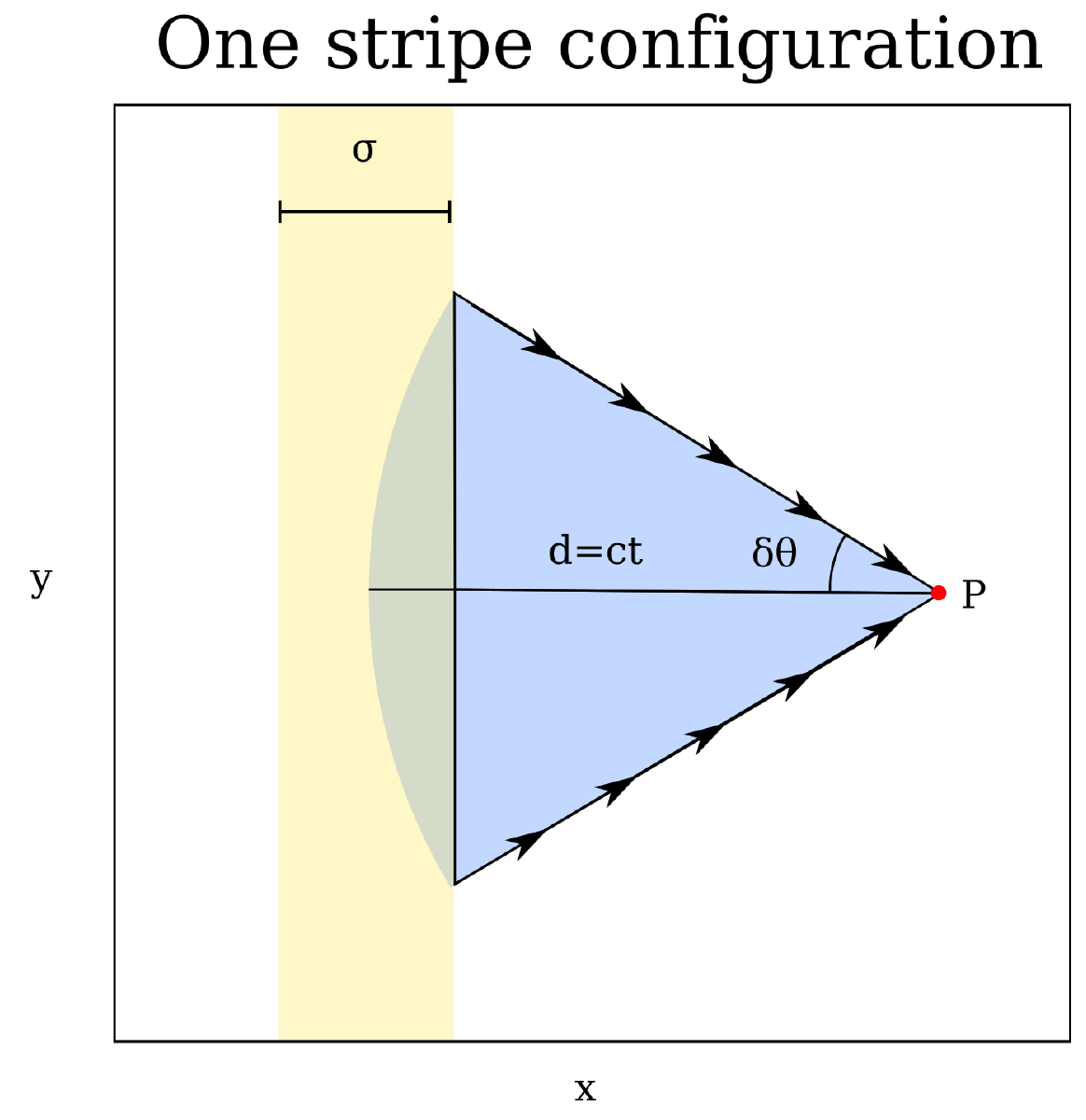}
\caption{{\it Left:} Schematic diagram of the box in the ``one dot configuration.''  The maximum width of the angular distribution $\delta\theta$ is determined by the dot size in the dot configuration. {\it Right:} Schematic diagram of the box in the ``one stripe configuration.'' Within the yellow stripe, ELN crossings are distributed in the plane following a Gaussian distribution of width $\sigma$. After a time $t$, the neutrino angular distribution in $P$ will have  maximum opening angle $\delta\theta$ because of  geometric effects. }
\label{fig:4}
\end{figure}

Before exploring the growth of  flavor instabilities in the different configurations assumed for our 2D box, we  adopt geometrical arguments to forecast how the angular distributions should evolve in the presence of advection. 
The left panel of Fig.~\ref{fig:4} shows a schematic diagram of the box in the ``one dot configuration.'' Let us consider a point $P =(x,y)$ outside the initial location of the dot ($|(x-x_{0})| > \sigma$). At  $t \sim (x-x_{0})/c$, neutrinos from various points across the dot and emitted along different $\theta$'s will reach $P$.  

It is easy to estimate the width of the angular distribution at any given time, which is dependent on the width of the initial distribution $b(\bar{b})$, see Eqs.~\ref{eq:gauss}-\ref{top2}. If the width of the initial  distribution is zero (i.e., we start with Dirac delta function), then at any given time $t$, only neutrinos traveling along a certain direction can reach the point $P$. In other words, for any point $P$ that was not on the dot initially, the angular distribution will still be a Dirac delta distribution in $\theta$.  
Understanding the limiting case with $b(\bar{b}) \rightarrow 0$ allows to draw insightful conclusions
regarding the evolution of the ELN crossings in time. 

Let us consider a point $P=(x,y)$  outside the ``dot'' in the 2D box, i.e.~$d^2=(x-x_{0})^{2} +  (y-y_{0})^{2} > \sigma^{2}$. From simple geometric considerations, it can be seen that the advective term acts like a narrow pass filter. In fact, at $t$ the (anti)neutrino angular distribution will have  width
\begin{eqnarray}
\delta \theta \approx \min\left[w, \arctan\left(\frac{\sigma}{d}\right)\right]\ ,
\end{eqnarray}
where $w$ is the width of the initial angular distribution of neutrinos or antineutrinos.

Similar considerations hold for the ``one stripe configuration,'' see the right panel of Fig.~\ref{fig:4}, the effect of advection is less prominent than that in the ``one dot configuration,'' but it exists nonetheless.
If the distance between $P$ and $x_{0}$ is $d$, then the width of the angular distribution at a certain time $t$ will be 
\begin{eqnarray}
\delta \theta \approx \min\left[w, \arctan\left(\sqrt{\frac{\sigma}{d}}\right)\right]\ ,
\label{deltheta}
\end{eqnarray} 
where $w$ is the width of the initial angular distribution, which is $b$ or $\bar{b}$ (Eqs.~\ref{top1} and \ref{top2}). Equation~\ref{deltheta} is  such that no matter how large $\sigma$ is, the angular width cannot be greater than $w$. The second argument of the $\min$ function in Eq.~\ref{deltheta} can be easily gleaned by noticing that the width of the angular distribution is given by the neutrino which is emitted at $x-x_{0}=\sigma$. 
If $\arctan\left(\sqrt{{\sigma}/{d}}\right) < w$, then the angular distributions of $\nu_{e}$ and $\bar{\nu}_{e}$ have the same width, and  fast conversions cannot occur. 

To gain physical intuition as of the meaning of Eq.~\ref{deltheta}, we  consider a limiting case where neutrinos have traveled a very large distance compared to the initial width of the ELN crossings i.e. when $\sigma/d \rightarrow 0$. In this limit, Eq.~\ref{deltheta} gives $\delta\theta \approx 0$ which corresponds to Dirac-delta's for the final angular distributions of (anti)neutrinos. As one can see in Fig.~\ref{fig:4}, in this limit the neutrinos reaching a fixed point $P$ after time $t$ are from the same direction, which results in final angular distributions with an almost vanishing $\delta\theta$. In other words, the finite size of the source cannot be resolved by an observer at infinite distance.

We now explore the role of the advective term in the neutrino equations of motion. 
We expect that the advective term will have  several effects on the neutrino flavor evolution. On the one hand, the advective term should diffuse  any eventual ELN excess localized in a small spatial region,  diluting it over a broader region. On the other hand, the parameter $\zeta$ should be modified  as a result of advection. Moreover, as it will become clear in the following, by modifying the instability parameter, the advective term will also affect the occurrence of fast conversions. In the non-linear regime, the effect of advection can be grasped by similar arguments. At a given time, neutrinos in a different phase of bipolar oscillations arrive at the  point $P=(x,y)$. This leads to  erasing the bipolar nature of oscillations, since oscillations with different phases overlap with each other.

The time-scale required for the advective term to wipe out the instability parameter ($\zeta \rightarrow 0$) is given by the time at which the two arguments of the min function in Eq.~\ref{deltheta} are comparable:
\begin{eqnarray}
t_{\textrm{conv}} \approx \frac{\sigma}{w^{2}}\ .
\label{tconv}
\end{eqnarray}
In this case, the width of the angular distribution of $\nu_{e}$ and $\bar{\nu}_{e}$ becomes independent of the initial angular distribution, and it is the same for $\nu_{e}$ and $\bar{\nu}_{e}$. 

The characteristic time scale of neutrino advection, $t_{\textrm{conv}}$, should be compared with the other characteristic time scale of the system  $t_{\textrm{osc}}$, which defines  the time required for  flavor transformations to occur. In fact, because of flavor instabilities, the off-diagonal term of the density matrix evolves like 
$ \rho_{ex}(t=0) \exp(\kappa t_{\textrm{osc}})  = \mathcal{O}(\rho_{ee})$;
 $t_{\textrm{osc}}$  obviously depends on the initial magnitude of the off-diagonal term, the effective neutrino number density, and  $\zeta$. In turn, the latter two evolve in time because of the advective term in the equations of motion.

\subsection{Flavor evolution in the one dot configuration}
\label{sec:dot}

 We now consider the case of neutrinos and antineutrinos initially localized within a dot in the 2D box (``one dot configuration''), see Eq.~\ref{eq:dot} and the left top panel of Fig.~\ref{fig:1}. 
We assume the initial angular distributions  as described in Sec.~\ref{sec:setup} and fix $b=\pi/6$ and $\bar{g}=0.5$ (see Sec.~\ref{sec:setup} for more details). Animations of
the temporal evolution of the   $\nu_e$ and $\bar\nu_e$ angular distributions, as well as the diagonal and off-diagonal terms of the neutrino and antineutrino density matrices are provided as \href{https://sid.erda.dk/share_redirect/BAdkN7XRul/index.html}{Supplemental Material}.

\begin{figure}[b!]
\includegraphics[width=0.99\textwidth]{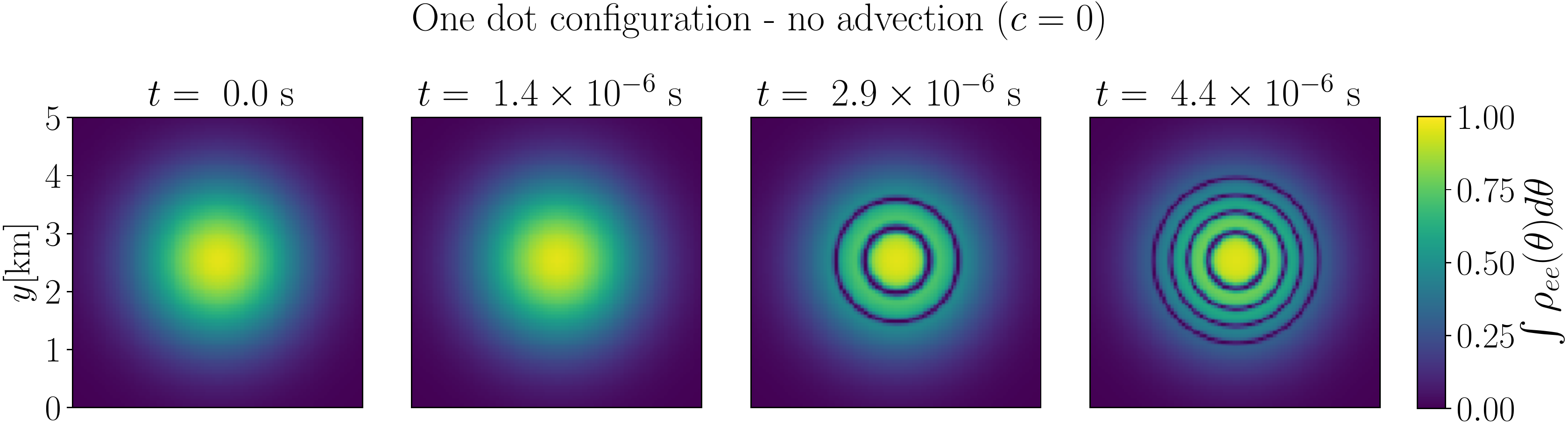}
\includegraphics[width=0.99\textwidth]{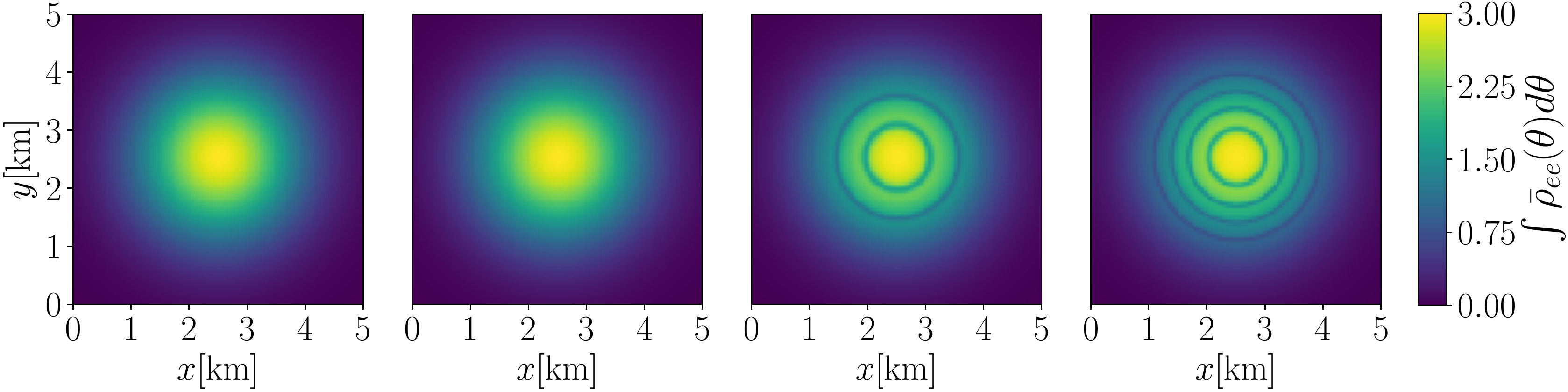}
\caption{Evolution of $\int d\theta \rho_{ee} d\theta$ (top) and $\int d\theta \bar\rho_{ee} d\theta$ (bottom)  in the 2D box for the ``one dot configuration'' (see Fig.~\ref{fig:1}) in the absence of advection. The four panels from left to right correspond to four different time snapshots ($t=0.0, 1.4 \times 10^{-6}$, $2.9 \times 10^{-6}$, and $4.4\times 10^{-6}$~s). The concentric circles in the neutrino density iso-contours highlight the regime of bipolar oscillations. The contours of $\bar{\nu}_{e}$ are less extreme than the ones of $\nu_e$, since flavor conversions occur in such a way to guarantee the lepton number conservation.
}
\label{fig:5}
\end{figure}
The top panels of Fig.~\ref{fig:5} show  iso-contours of $\int \rho_{ee}(\theta) d\theta$ for four different time snapshots, from left to right respectively, in the absence of advection. Bipolar oscillations develop over time and give rise to concentric circles in the neutrino density iso-contours (see $t=2.9, 4.4 \times 10^{-6}$~s). This behavior is identical to what is shown in Fig.~\ref{fig:2} for a selected point in the $(x,y)$ plane. The bottom panels of Fig.~\ref{fig:5} show the correspondent evolution of $\bar{\nu}_{e}$  for the same time snapshots, highlighting  the collective behavior of neutrinos and anti-neutrinos. It should be noted that the contours of $\bar{\nu}_{e}$ are less extreme than the ones of $\nu_e$, since $\nu_e$ and $\bar\nu_e$ can transform as long as the lepton number is conserved. 

\begin{figure}
\includegraphics[width=0.99\textwidth]{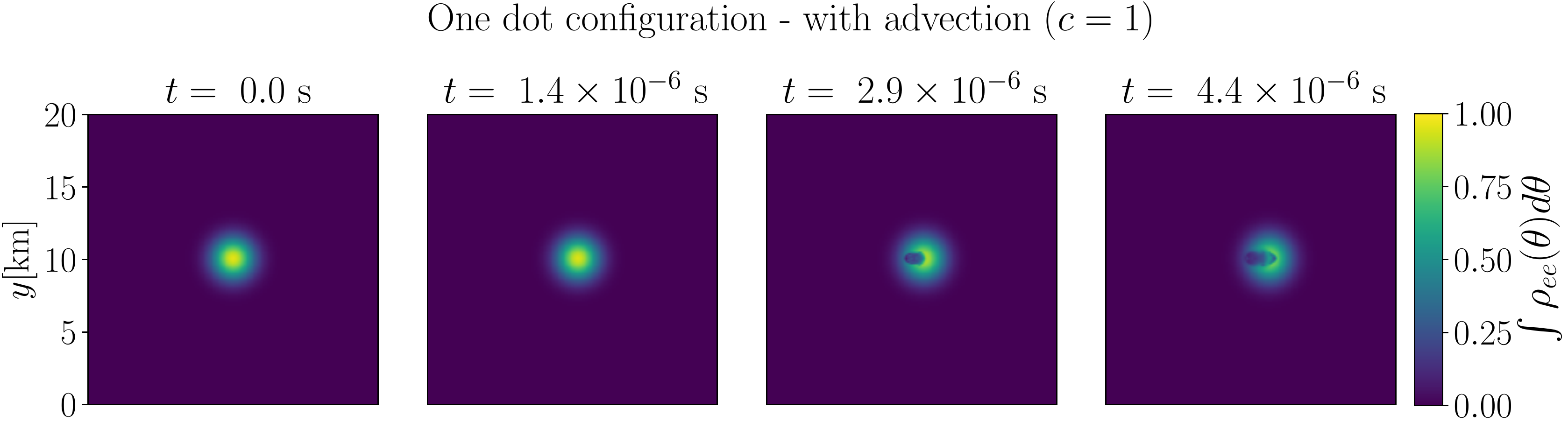}
\includegraphics[width=0.99\textwidth]{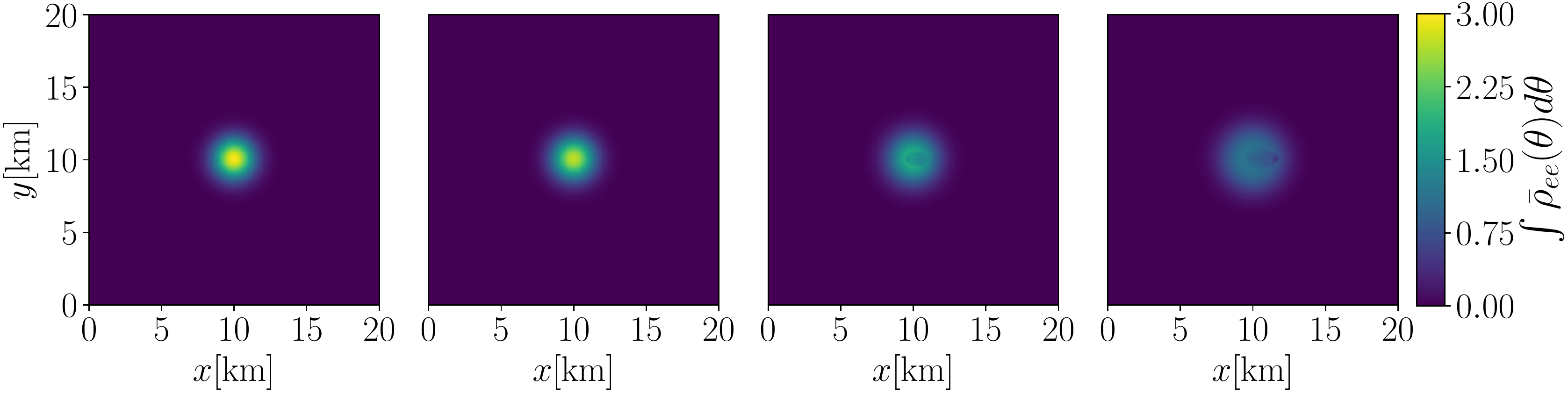}
\includegraphics[width=0.91\textwidth]{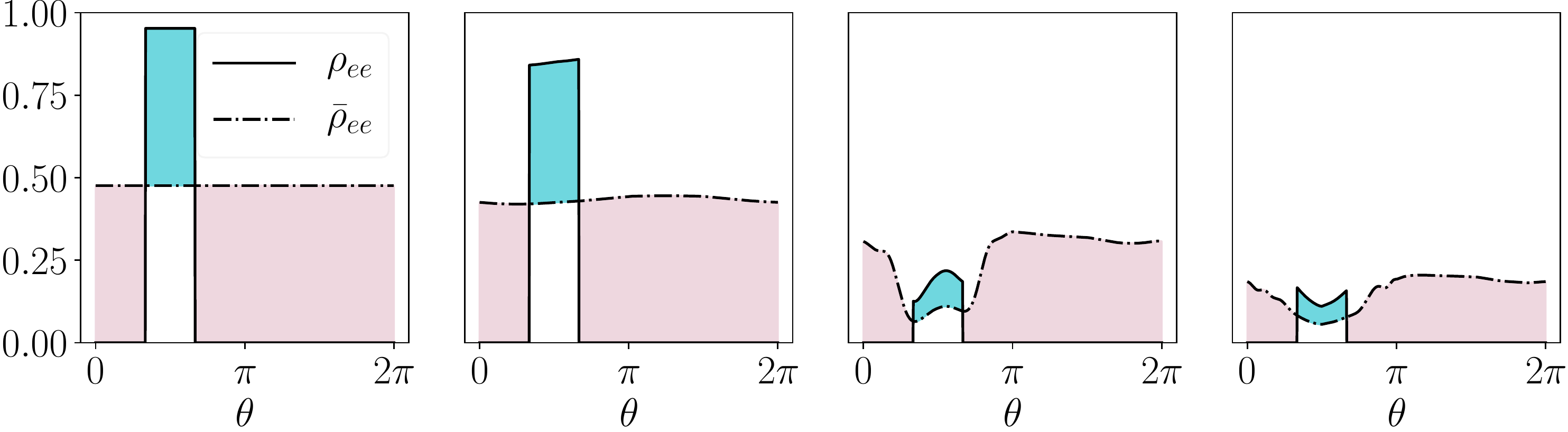}
\caption{
{\it Top and middle:} Same as Fig.~\ref{fig:5}, but in the presence of advection.  Because of advection  flavor transformations occur in a localized region and the bipolar structure is smoothed out. 
 Since the angular distribution of  $\int \bar{\rho}_{ee} d\theta$ is isotropic, advection favors a decrease of the local density of $\bar\nu_e$  very rapidly (see bottom panel of Fig.~\ref{fig:5} for comparison). 
{{\it Bottom:} Evolution of the angular distributions of $\nu_{e}$ and $\bar{\nu}_{e}$ as a function of $\theta$ for a comoving  point in our 2D box. Due to advection, the  $\zeta$ parameter  decreases with time, hindering fast conversions.}
}
\label{fig:6}
\end{figure}

The picture described above is drastically modified by advection, as shown in Fig.~\ref{fig:6}. By comparing this figure to Fig.~\ref{fig:5} (i.e., to the case without advection), one notices that advection contributes to erase the small scale structures characterizing flavor conversions: the bipolar behavior of flavor conversions disappears and  the flavor change is confined to a more localized region.  The correspondent reduction in the value of the ELN parameter $\zeta$ can be clearly seen in the bottom panels where the $\nu_e$ and $\bar\nu_e$ angular distributions are shown.  The reduction of the ELN parameter $\zeta$, in turn, drives the system towards a more stable configuration disfavoring further flavor change.

\begin{figure}
\begin{center}
\includegraphics[width=0.7\textwidth]{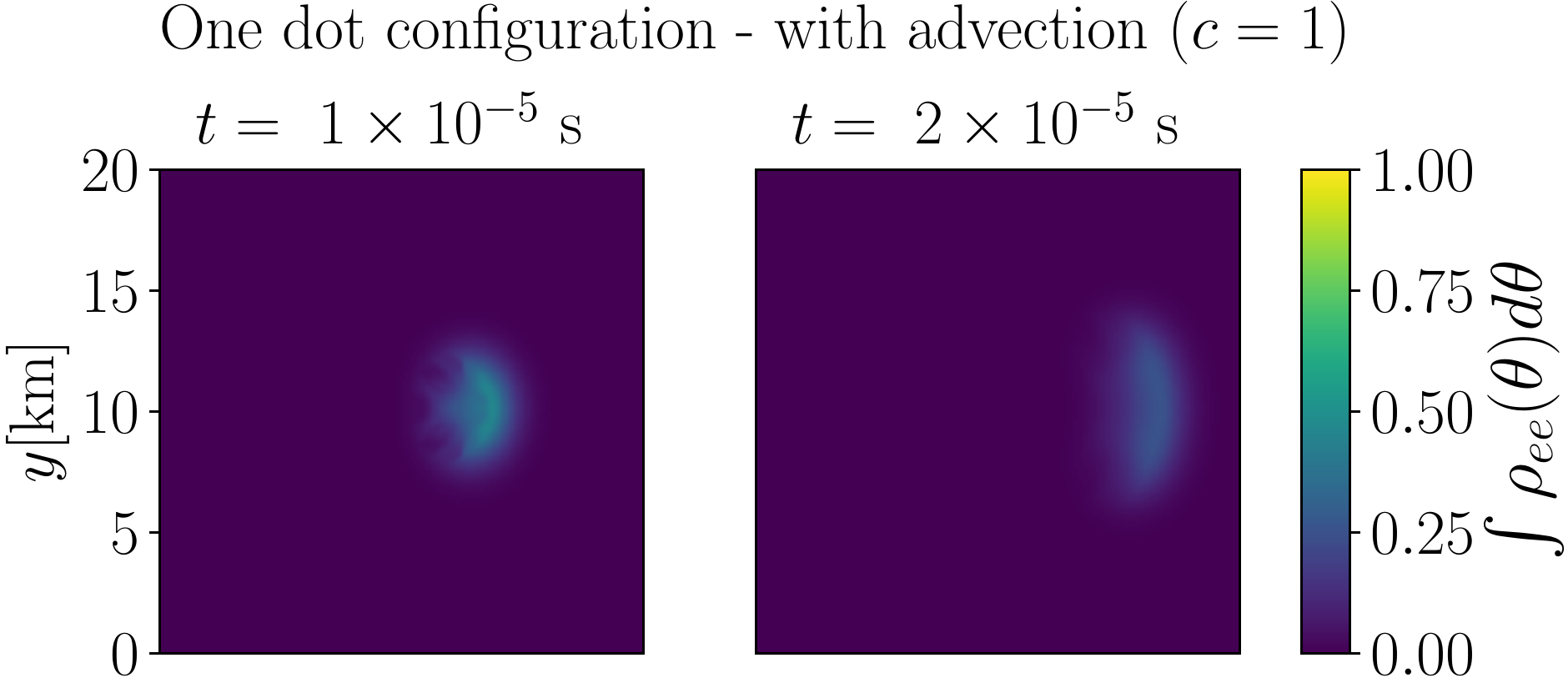}\\
\vspace{0.25cm}
\includegraphics[width=0.7\textwidth]{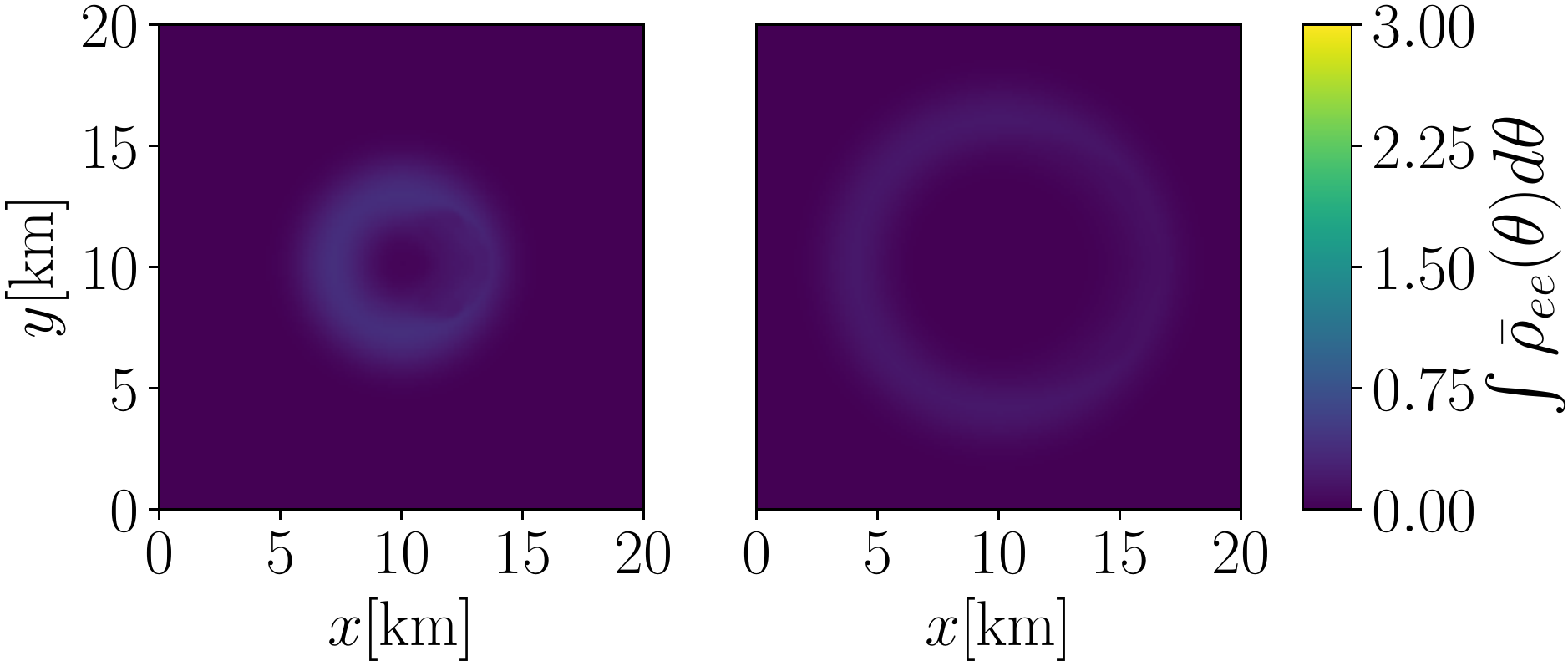}
\end{center}
\caption{
Same as in the top and middle panels of Fig.~\ref{fig:6}, but for $t=10^{-5}$ and $2\times 10^{-5}$~s. The initial angular distributions of $\nu_e$'s are forward peaked, hence neutrinos  tend to diffuse forward because of  non-negligible advection. This effect becomes even more extreme for $\bar\nu_e$'s that have an initially homogeneous  angular distribution. Advection  also contributes to progressively reduce the local density in time, spreading all particles throughout the box.
}
\label{fig:7}
\end{figure}

As time progresses, the $\zeta$ parameter is reduced so much that diffusion becomes the only dominant phenomenon as shown in Fig.~\ref{fig:7}, where the iso-contours of $\int \rho_{ee}(\theta)d\theta$  and $\int \bar{\rho}_{ee}(\theta)d\theta$ are displayed in our 2D box for later times ($t=10^{-5}$ and  $2\times 10^{-5}$~s). Since the initial $\nu_e$ angular distribution is forward peaked,  $\nu_e$'s tend to diffuse forward more prominently; while, the effect of advection is stronger on $\bar\nu_e$'s that have an initially homogeneous  angular distribution.

\begin{figure}
\begin{center}
\includegraphics[width=0.7\textwidth]{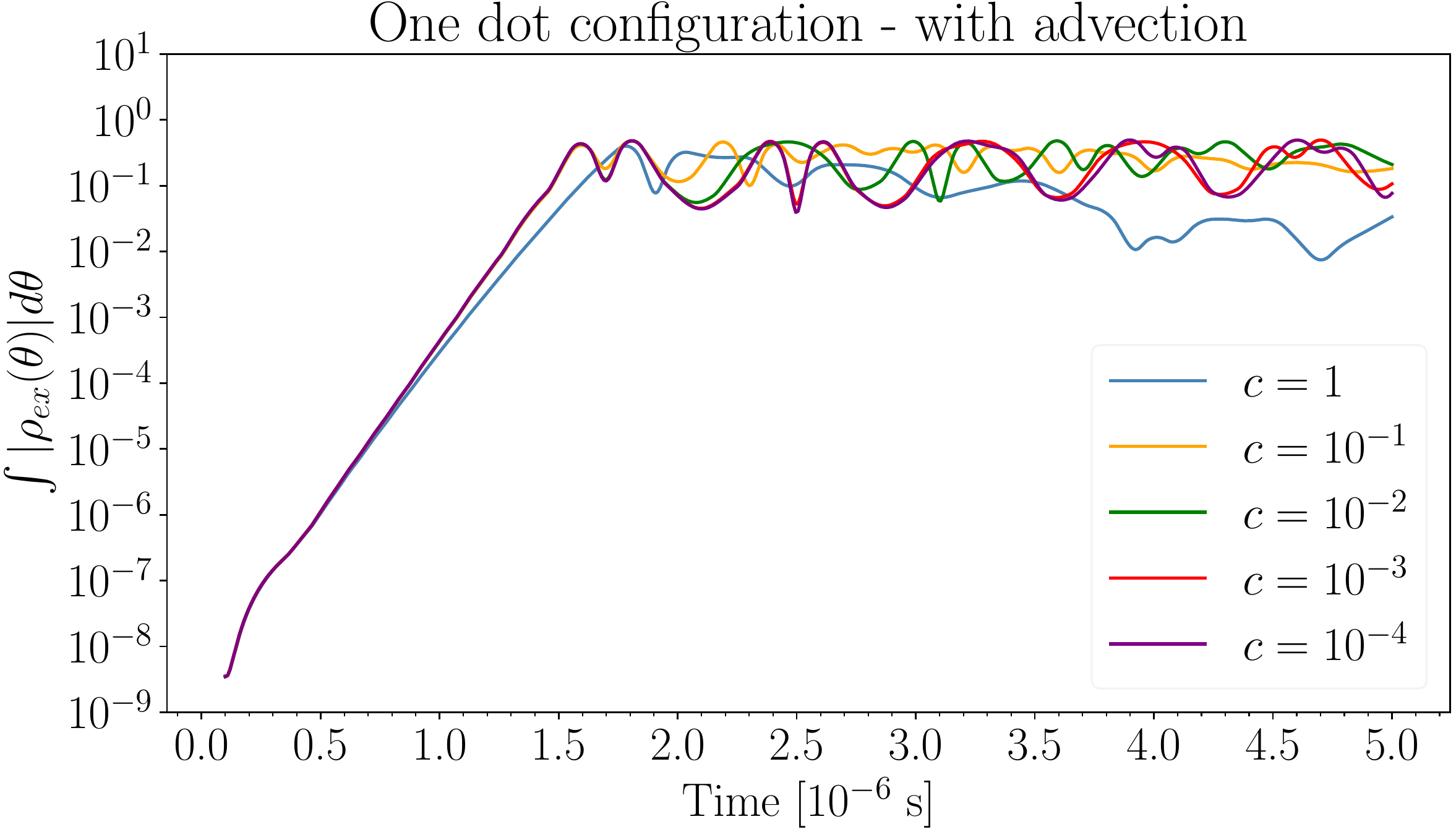} 
\includegraphics[width=0.7\textwidth]{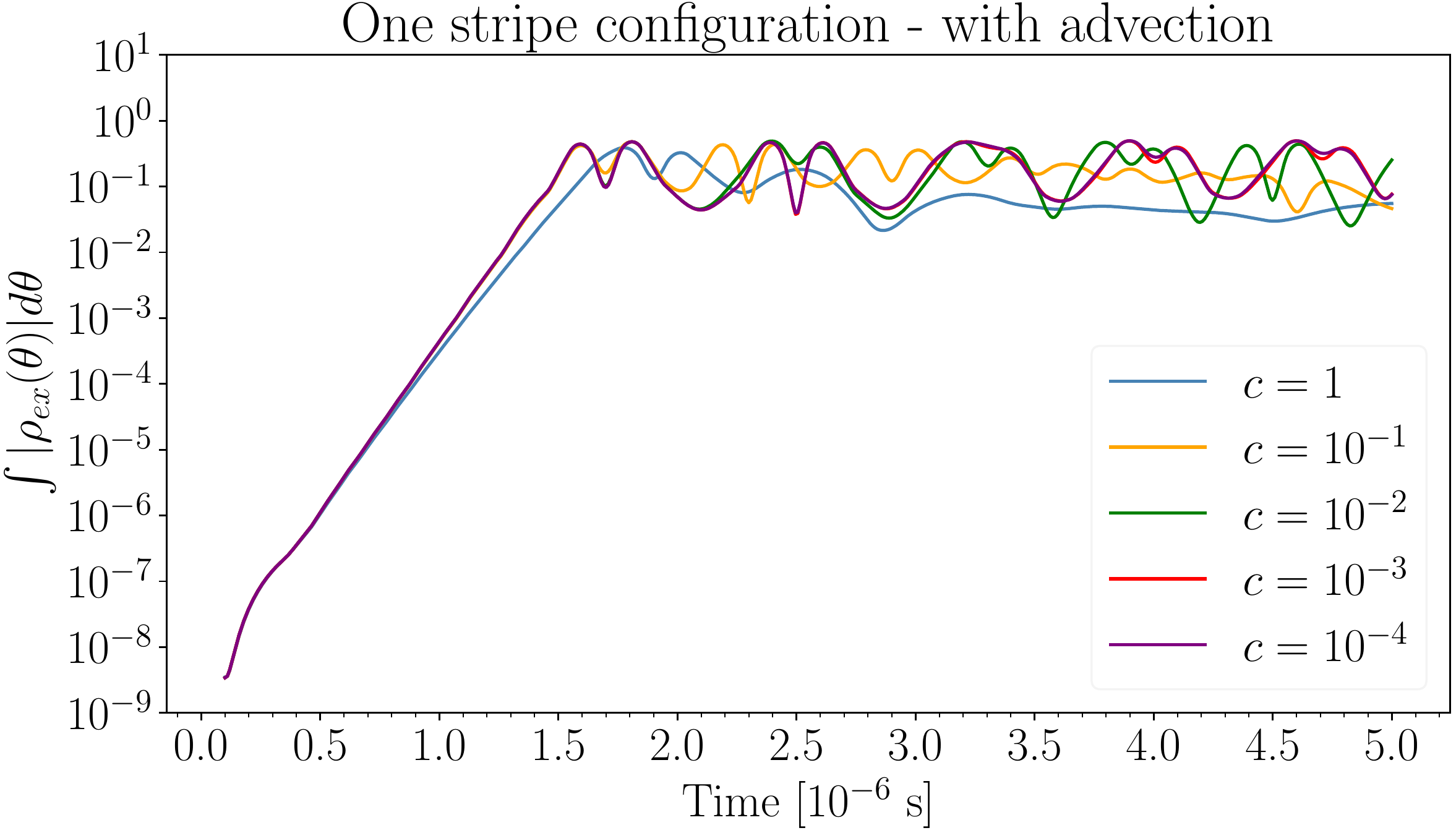}
\end{center}
\caption{Temporal evolution of $\int |\rho_{ex}(\theta)| d\theta$ for different values of the advective velocity $c$ for a comoving point  in the 2D box for the  ``one dot configuration'' (top) and for the  ``one stripe configuration'' (bottom). For small values of $c$ the deviation from the bipolar oscillations is minimal. As $c$ increases the $\zeta$ parameter becomes smaller, and the overall flavor conversion probability tends to reach a smaller asymptotic value. Because of the different geometry, the impact of advection is slightly less pronounced in the ``one stripe configuration.''
}
\label{fig:8}
\end{figure}

In order to highlight how the bipolar regime is modified by advection, the top panel of Fig.~\ref{fig:8} shows the evolution of $\int |\rho_{ex}(\theta)| d\theta$  for a range of values of  $c$ different from our  default value ($c=1$) for a comoving point in the 2D box. Although this is an academic exercise, it shows that  advection becomes more and more relevant as $c \rightarrow 1$. In addition, as $c$ increases, the $\zeta$ parameter is also affected since the ELN crossings are smeared and the overall flavor conversion probability tends to reach a smaller asymptotic value.  

The ``one dot configuration'' mimics what would happen in SNe or compact binary mergers in the presence of localized  ELN excess, e.g.~generated by stochastic hydrodynamical fluctuations. We can conclude that advection smears the ELN crossings in this configuration most likely leading to minimal changes in the flavor configuration.

\subsection{Flavor evolution in the one stripe configuration}\label{sec:onestripe}

We  now focus on the ``one stripe configuration'' of our 2D box, see the  top panel on the right of Fig.~\ref{fig:1}.  
Similarly to the ``one dot configuration,'' we  investigate the flavor evolution, first in the absence of advection and then by including the advective term in the neutrino equations of motion for  $b=\pi/6$ and $\bar{g}=0.5$ (see Sec.~\ref{sec:setup} for more details). We provide animations of  
 $\nu_e$ and $\bar\nu_e$ angular distributions, the diagonal and off-diagonal terms of the neutrino and antineutrino density matrices as \href{https://sid.erda.dk/share_redirect/BAdkN7XRul/index.html}{Supplemental Material}.

\begin{figure}
\includegraphics[width=0.99\textwidth]{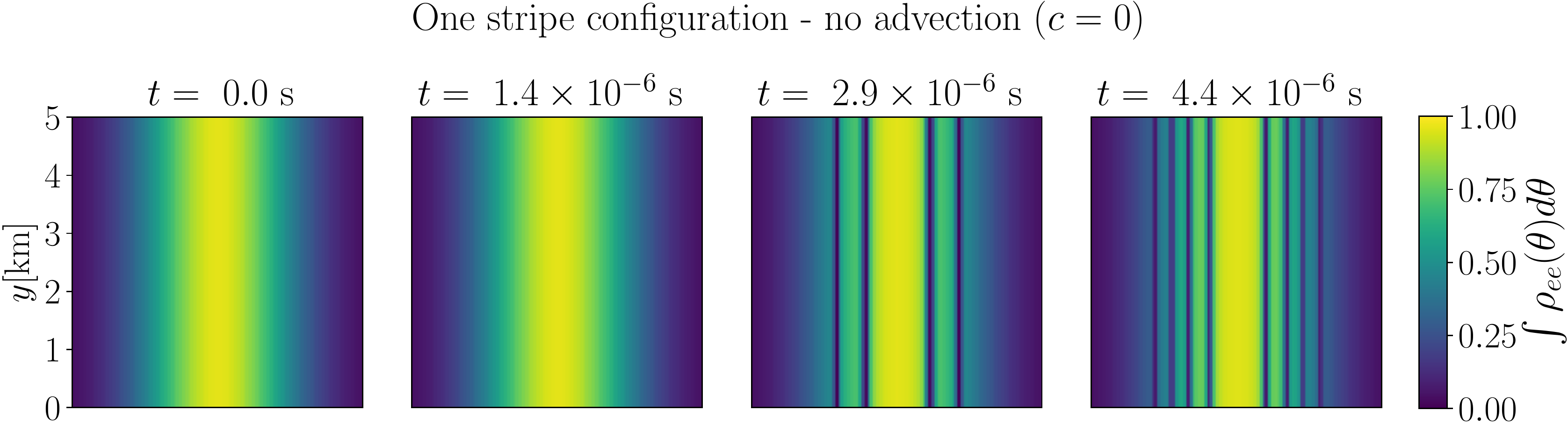}
\includegraphics[width=0.99\textwidth]{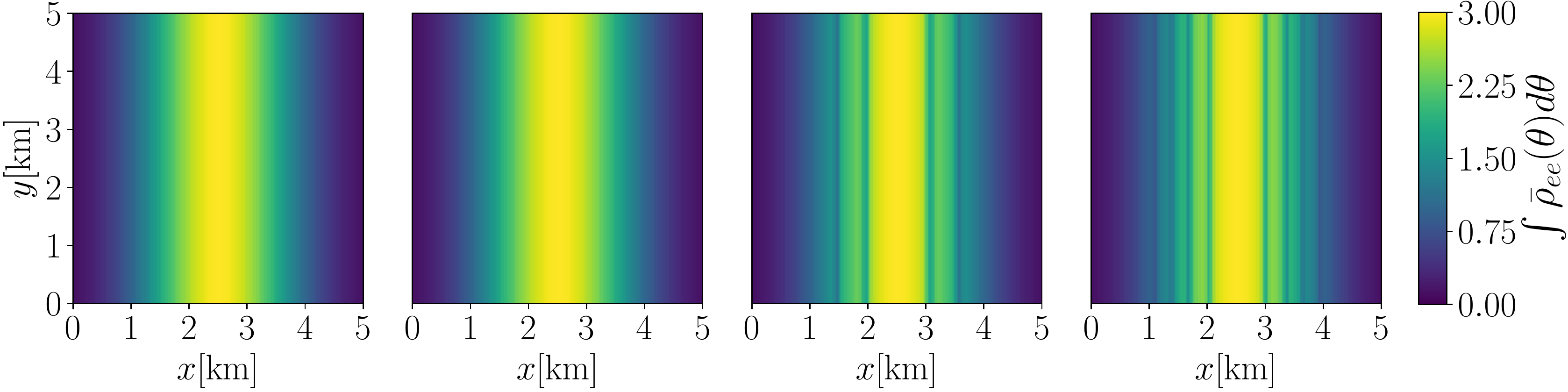}
\caption{Same as Fig.~\ref{fig:5}, but for the ``one stripe configuration.'' Neutrino-neutrino interactions lead to a bipolar structure in the flavor evolution in the non-linear regime.}
\label{fig:9}
\end{figure}
\begin{figure}
\includegraphics[width=0.99\textwidth]{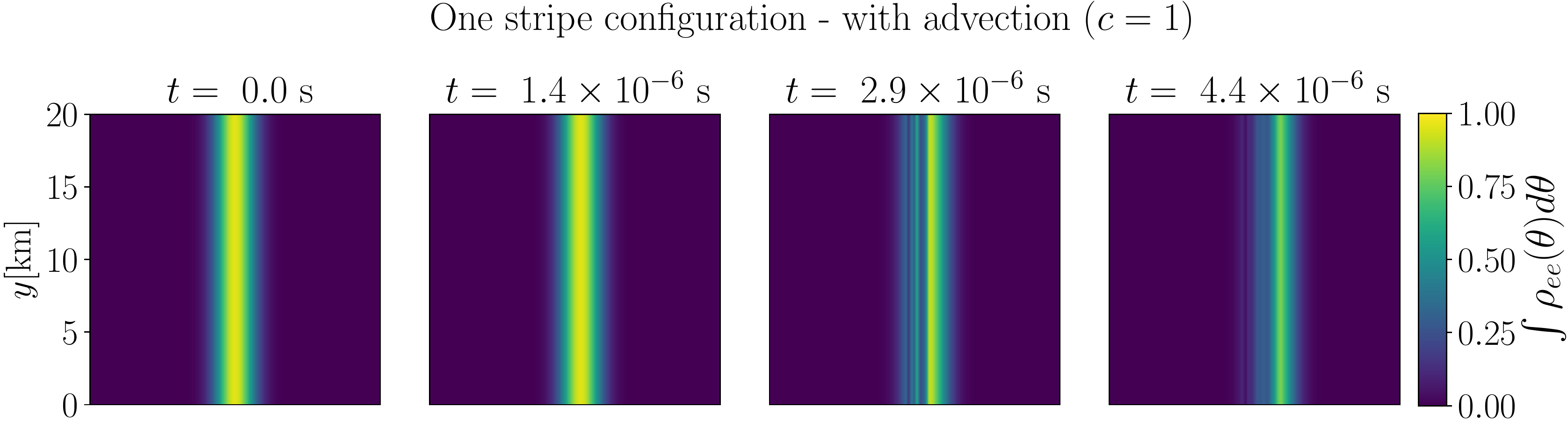}
\includegraphics[width=0.99\textwidth]{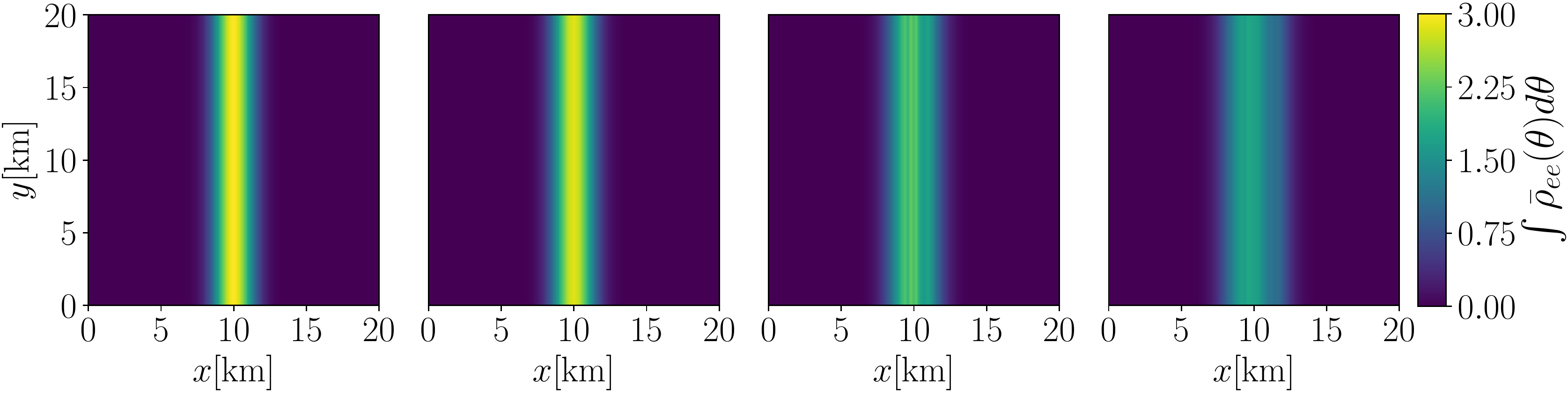}
\includegraphics[width=0.91\textwidth]{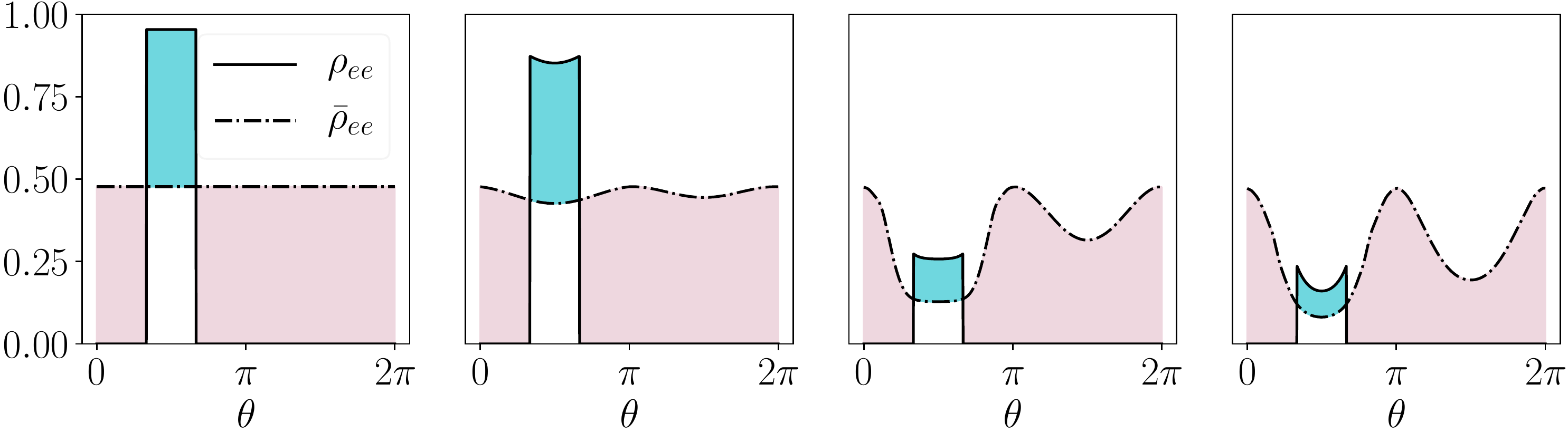}
\caption{Same as Fig.~\ref{fig:6}, but for the ``one stripe configuration.'' The ELN parameter becomes smaller as a function of time. However, its decrease is slower than in the  ``one dot configuration'' because of the differences in the initial geometry of the two configurations.}
\label{fig:10}
\end{figure}

In the absence of advection,  the bipolar nature of flavor conversions is evident also for the ``one stripe configuration''  as shown  in Fig.~\ref{fig:9} where four snapshots of the iso-contours of $\int \rho_{ee}d\theta$ and  $\int \bar\rho_{ee}d\theta$ are shown. 
For comparison, the effect of advection is visible in Fig.~\ref{fig:10}. As evident from the bottom panels, the $\zeta$ parameter becomes smaller as a function of time. However, in comparison to the ``one dot configuration,'' the ELN parameter $\zeta$ decreases more slowly because of the differences in the initial geometry between the two configurations.  

\begin{figure}
\begin{center}
\includegraphics[width=0.7\textwidth]{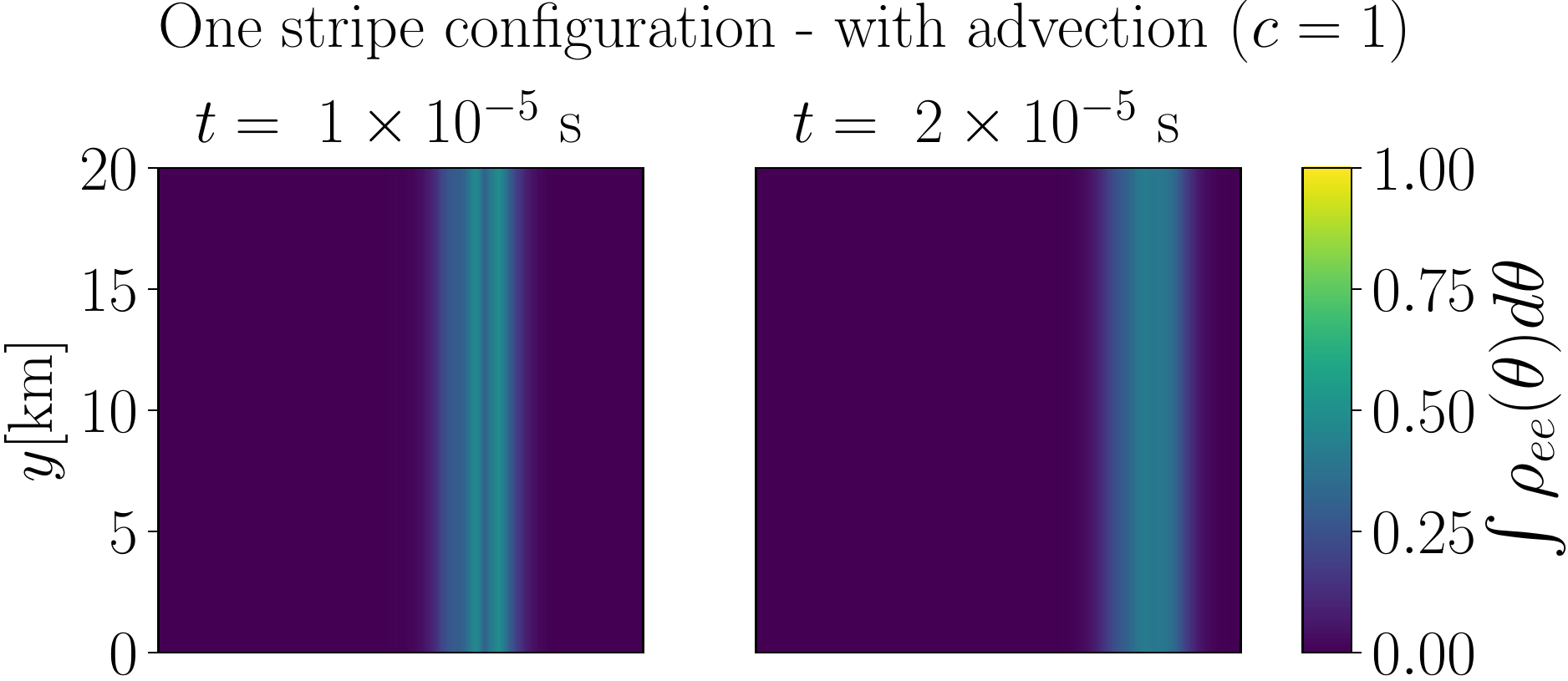}\\
\vspace{0.25cm}
\includegraphics[width=0.7\textwidth]{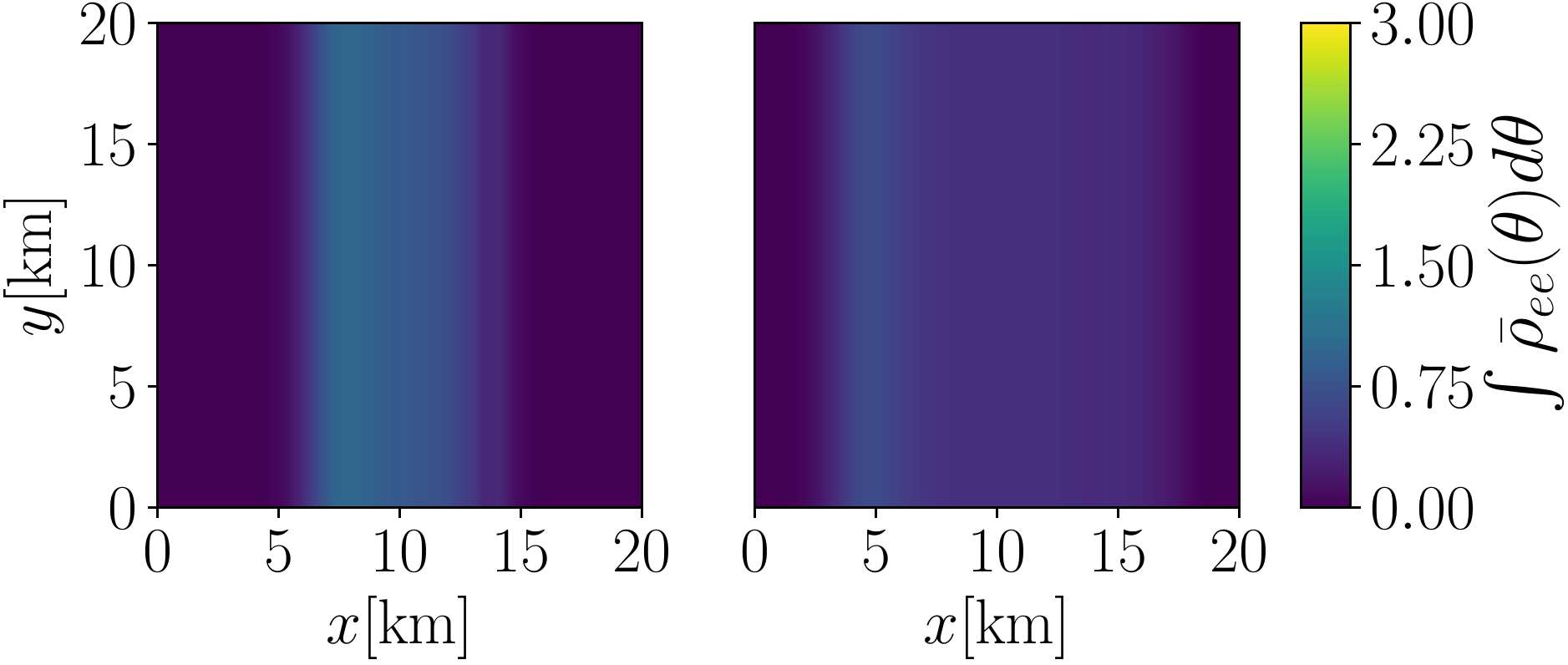}
\end{center}
\caption{Same as Fig.~\ref{fig:7}, but for the ``one stripe configuration.'' Advection spreads the original stripe across our 2D box, disfavoring the occurrence of favorable conditions for fast conversions.}
\label{fig:11}
\end{figure}

The effect of advection is less pronounced  in the ``one stripe configuration'' because the neutrinos reaching any spatial point $(x,y)$ in our 2D box are emitted from a region more extended in space than the ``one dot configuration'' and therefore  have a larger spread in their angular distributions. As a consequence, longer time is needed  to obtain  completely forward peaked angular distributions. This can be also seen in Fig.~\ref{fig:11} which focuses on the flavor evolution at later times ($t=1 \times 10^{-5}$ and $2 \times 10^{-5}$~s).

The effect of advection on $\int \rho_{ex} d\theta$ in the ``one stripe configuration'' is shown for a comoving point in the bottom panel of Fig.~\ref{fig:8} for various values of $c$. For $c \rightarrow 1$,  $\int \rho_{ex} d\theta$ is closer to unity  than in the ``one dot configuration'' (see top panel of Fig.~\ref{fig:8} for comparison).

This configuration of the 2D box tends to mimic what would happen in a SN patch in the presence of a front of ELN crossings. As one can see, unless the ELN crossings are self-sustained (as it could be in the case of LESA), they would be wiped out by the neutrino advective term. As a consequence, fast pairwise conversions would only lead to  partial flavor conversion. Notably, our setup overestimates the effect of flavor conversions since we assume $\mu=\mathrm{const.}$ in the box and maximize the initial $\zeta$ by assuming an isotropic distribution for $\bar\nu_e$. In a realistic case, $\mu$ would tend to decrease as the distance from the decoupling region increases and $\zeta(t=0~\mathrm{s})$ would be likely smaller than in our case given that the angular distribution of $\bar{\nu}_e$ is not isotropic outside the proto-neutron star radius.

\section{Outlook and conclusions}
\label{sec:conc}

Compact astrophysical objects are so dense in neutrinos that  quantum effects are expected to manifest at macroscopic scales. In this work, we explore  an interesting and insightful interplay between the neutrino pairwise conversions (quantum effect) and the propagation of the neutrino field driven by the advective term in the equations of motion (classical effect). In order to do this, for the first time, we track the neutrino flavor evolution within a (2+1+1) framework, i.e.~we solve the neutrino equations of motion in time, two spatial dimensions, and one angular variable.

We explore a simplified scenario with constant neutrino--neutrino potential for the sake of simplicity, however we mimic configurations similar to the ones that could occur in compact astrophysical objects where favorable conditions for fast pairwise conversions  have been found through the stability analysis. In particular, we explore two different configurations: 1.   one  localized excess of particles with  electron lepton number (ELN) crossings, mimicking ELN fluctuations  that could occur because of stochastic hydrodynamical fluctuations; 2. neutrinos and antineutrinos initially localized along one stripe in our 2D box,  mimicking a situation similar to LESA.

We generalize the conditions leading to the development of fast pairwise conversions introducing the instability parameter $\zeta$ that broadly captures the essence of the shape of the neutrino angular distributions leading to flavor instabilities. The instability parameter $\zeta$ along with the effective strength of  neutrino-neutrino interactions determines the instantaneous growth rate of the off-diagonal components of the density matrices. 
However, the numerical solution of the neutrino equations of motion highlights a fascinating interplay between the growth of fast pairwise conversions and neutrino advection. Our sophisticated numerical simulations show that the advective term in the equations of motion  hinders the growth of flavor instabilities, unless the front of ELN crossings (or  of neutrinos moving in the opposite direction of $\vec{v}$) is self-sustained in time.  

As a consequence, our simple model predicts that significant flavor evolution due to fast pairwise conversions can occur in the presence of the LESA instability (scenario 2), but would not be significant for a localized ELN excess (scenario 1). 
However, a more in-depth analysis including collisions is  mandatory, since collisions may damp the growth of flavor instabilities even in the presence of self-sustained crossings.

This work  demonstrates a critical limitation of  the linear stability analysis widely used in the field of collective neutrino conversions, as the  time scales that are relevant from the point of view of classical evolution (i.e., advection) may be comparable to the time scale of flavor conversions. In addition, the advective term in the equations of motion is such that the conditions leading to the growth of flavor instabilities are   dynamically affected from the surroundings. This aspect is not captured by the stability analysis.

Ours is the first numerical solution of the neutrino flavor evolution within a  sophisticated and dynamical multi-dimensional framework. Although our work is in no way the final setup resembling the evolution of the neutrino field in core-collapse supernovae or compact binary mergers, it  highlights the dynamical nature of flavor evolution.

\acknowledgments
We are grateful to the Villum Foundation (Project No.~13164), the Danmarks Frie Forskningsfonds (Project No.~8049-00038B), the Knud H\o jgaard Foundation, and the Deutsche Forschungsgemeinschaft through Sonderforschungbereich
SFB~1258 ``Neutrinos and Dark Matter in Astro- and
Particle Physics'' (NDM).

\bibliographystyle{JHEP}
\bibliography{convection1}
\end{document}